# Efficient and accurate simulation of vitrification in multi-component metallic liquids with neural-network potentials

Rui Su[1,*], Jieyi Yu[1], Pengfei Guan[2,1,*], Weihua Wang[3]

[1]Institute of Advanced Magnetic Materials, College of Materials & Environmental Engineering, Hangzhou Dianzi University, Hangzhou 310018, P. R. China

[2]Beijing Computational Science Research Center, Beijing 100193, P. R. China

[3]Songshan Lake Materials Laboratory, Dongguan 523808, China

***Corresponding authors:**

R. Su (surui@hdu.edu.cn)

P. F. Guan (pguan@csrc.ac.cn)

## Abstract

Constructing accurate interatomic potential and overcoming the exponential growth of structural equilibration time are challenges to the atomistic investigations of the composition-dependent structure and dynamics during the vitrification process of deeply supercooled multi-component metallic liquids. In this work, we describe a state-of-the-art strategy to address these challenges simultaneously. In the case of the representative Zr-Cu-Al system, in combination with a general algorithm for generating the neural-network potentials (NNP) of multi-component metallic glasses effectively and accurately, we propose a highly efficient atom-swapping hybrid Monte Carlo (SHMC) algorithm for accelerating the thermodynamic equilibration of deeply supercooled liquids. Extensive calculations demonstrate that the newly developed NNP faithfully reproduces the phase stabilities and structural characteristics obtained from the *ab initio* calculations and experiments. In the combined NNP-SHMC algorithm, the structure equilibration time in the deeply supercooled temperatures is accelerated by at least five orders of magnitudes, and the quenched glassy samples exhibit comparable stability to those prepared in the laboratory. Our results pave the way for the next-generation studies of the vitrification process and, thereby the composition-dependent glass-forming ability and physical properties of multi-component metallic glasses.

# Introduction

Metallic glasses (MGs) have shown great potential for a broad range of applications as structural[1,2] or functional[3,4] materials. Like other glasses, MGs are often fabricated by rapidly quenching the high-temperature melts that bypass crystallization[5]. However, the atomic-level process of the vitrification of MGs from their parent supercooled liquids is far from clear. So far, many experiments have shown that the two-body structural correlations show no qualitative changes during the vitrification process[6]. Due to the limited spatial and temporal resolutions of current experimental techniques, revealing the microscopic structural motifs in metallic glasses and their evolution during vitrification still faces significant challenges[7–9]. In contrast, computer simulations[10–14] provide effective ways to explore the atomic-level structure evolution during the vitrification process. However, two fundamental challenges of the current computer simulations exist: firstly, the complex atom interactions in multi-component MGs limit the accuracy of the over-simplified embedded-atom method (EAM) based inter-atomic potentials; secondly, the exponential growing structure equilibration time near the glass transition temperature strongly hinders the preparation of annealed MG samples comparable to the laboratory-made ones. Only by solving these two problems simultaneously, it is possible to reveal the underlying mechanisms behind the unique physical properties of amorphous alloys, which are strongly tied to their chemical compositions and thermal histories.

The local chemical bonding can be complicated in realistic MGs, which are frequently made up of multiple metal- or metalloid- elements combinations. To adequately describe the intricate atom interactions in such realistic MGs systems, *ab initio* molecular dynamics (AIMD) has been widely used. However, the small timescale ($\sim ps$ or quenching rate $Q > 10^{13}\ K/s$) and system size (~100 atoms) introduce significant statistical fluctuations in the calculation results and hinder the development of medium-range structural order in MGs. Additionally, constant pressure (*NpT*) simulation of MGs is expensive due to the requirement of a large basis cutoff to counteract the artificial Pulay stresses[15]. Thus, people often resort to classical molecular dynamics simulations with the embedded-atom method (EAM) potentials. While the longer ($10^2 - 10^4\ ns$) and larger ($10^3 - 10^6$ atoms) simulations of MGs become possible for EAM-based MD simulations, the reliability of EAM is crucial for comparing the simulation results with those from experiments. For instance, researchers recently found that the EAM potential of the binary Fe-P system, which has been used in several simulations of the cavitation processes in the MGs[16], undergoes room-temperature spinodal decomposition that significantly affects its fracture behaviors[17]. Another example is that several EAM potentials of the Cu-Zr binary system overestimate the phase stability of the Laves phases, which strongly encourages the formation of the Laves phase in the undercooled Cu-Zr liquids[18]. The inaccuracies of EAM potential come from two sources: firstly, EAM potentials are usually fitted against a small set of *ab initio* calculated configurations, which must be carefully selected to cover the representative bonding environments; secondly, the EAM model does not describe the angular interactions well, which is essential for MGs containing metalloid or non-metal elements.

Machine learning potentials[19–21] (MLPs), especially neural-network potentials (NNPs) trained on extensive *ab initio* datasets have been applied to MGs, which allows faster and larger MD simulations to be performed at the near-*ab initio* accuracy. As *there is no free lunch*, the transferability of MLPs strongly depends on the diversity of *ab initio* training database, and the computational complexity is often significantly greater than that of the EAM potentials, which makes it a challenging task to apply MLPs on the studying of the ultra-slow vitrification process of the glass-forming liquids.

Until recently, the swap Monte Carlo (MC) method[22] has been proven to obtaining very stable glass samples using simple pair potentials. For example, stable "metallic glasses" samples have been obtained by the simple pair Lennard-Jones (LJ) potentials[23,24] with swap MC. However, simple LJ potentials are clearly far from experimental measurements as the actual chemical bonding can be rather complex which usually contains directional interactions that cannot be described by the over-simplified pair potentials.

In this work, we present novel and effective algorithms to overcome the accuracy and time-scale issues when simulating the vitrification process in multi-component metallic glass-forming liquids. To this end, we first describe a novel structure descriptor based on the rigorous decomposition of local atom environments, which allows complete and computationally efficient mathematical representations of local structures. Then, a general two-stage workflow is suggested for constructing an extensive *ab initio* training database that effectively covers the chemical and structural space of the underlying alloy system. Finally, the novel atom Swapping assisted Hybrid Monte Carlo (SHMC) is presented to accelerate the equilibration of deeply supercooled liquids, which is thoroughly optimized for effective working with NNPs. Taking the well-known Zr-Cu-Al ternary MGs as a representative, we construct Zr-Cu-Al NNP that predicts crystalline phase stabilities and structure correlation functions in excellent agreement with *ab initio* and experimental results. Our performance analysis shows that our NNP implementation is at least one order of magnitude faster than the previous NNP implementation. Amazingly, when combining NNP with our SHMC method, we accelerate the equilibrium of the deeply supercooled $Zr_{46}Cu_{46}Al_8$ liquid by five orders of magnitudes, which is only ~30 K above the glass transition temperature. Further, SHMC quenching quickly brings the liquid to the glass state that is as stable as the laboratory-made MGs, which is firstly reported for NNP-based simulations. Thus, our work provides a comprehensive and general computational framework to understand the structural and chemical evolution in the vitrification process at an accuracy level close to the *ab initio* simulations, which hopefully pave the way for next-generation simulations of physical properties of compositionally complex metallic alloys.

# Results

### Neural network inter-atomic potential

Neural network inter-atomic potential is a pure mathematical potential model that maps the local energy and forces of each atom to its local chemical environment. To this end, fixed-length descriptor vectors are calculated as functionals of the per-atom local chemical environments, and the local energy and force of each atom are computed by passing the descriptors through the feed-forward neural networks. On the space of structural descriptors that NNP was trained, the structure energy and forces can be accurately predicted as the reference density functional theory (DFT) calculations, but with significantly less computation effort. In recent years, various NNP implementations have been reported for different chemical systems[20,25,26]. Here, we chose to develop the specific software implementation of NNP that we called NNAP (short for Neural Network inter-Atomic Potential), which is designed from scratch to be user-friendly and specially optimized for describing the multi-component MGs.

Like most NNPs, the potential energy of an atom configuration $\sigma$ in the NNAP implementation is expressed as the summation of per-atom energies:

$$E(\sigma) = \sum_{i \leq n} E_{atom}(\sigma_i^{R_c}) \tag{1}$$

where $\sigma_i^{R_c}$ defines a local environment of atom $i$ within a radial cutoff of $R_c$.

The local environment $\sigma_i^{R_c}$ itself, however, cannot be directly used as the input layer of the neural network since it does not have a fixed length. It is, therefore, necessary to have a structure descriptor that converts $\sigma_i^{R_c}$ to a fixed-length vector is required. Additionally, the functional form of the structure descriptor must ensure that the atomic energy is invariant under rotations, translations, and atom permutations.

While several kinds of structure descriptors have been proposed in the literature[20,27–30], in this work, we employ a self-developed one named "Spherical Chebyshev" (SC) basis in our NNP implementation. The reason for developing a new structure descriptor is based on the practical needs for simulating multi-component metallic glasses: firstly, the local structure motifs in the glass state are more complex than the ordered crystalline phases, which hinders the tuning of descriptors that require many parameters like the Behler-Parrinello (BP) descriptor[25]; secondly, simulating the vitrification process of MGs usually requires long-time quenching of the sample, which requires a high-performance numerical implementation. To be more specific, we want our new basis to have the following properties: firstly, the new descriptor should have a few systematically tunable parameters, such as the SOAP[30,31] (soft overlap of atom positions) descriptor to reduce the cost of the parameter tuning work; secondly, the math form should be simple for efficient numerical implementations; finally, the size of descriptor should be independent on the number of chemical species, which is critical for simulations of multi-component MGs since most of them contain more than three species.

To introduce the SC basis, we first consider the weighted and truncated local atom density function:

$$\rho(\boldsymbol{r}) = \sum_{i \neq j} w(t_j)\delta(\boldsymbol{r} - \boldsymbol{r}_{ij})f_c(\boldsymbol{r}_{ij}) \tag{2}$$

where $w(t_j)$ is a weight parameter of chemical specie $t_j$, $\delta(\boldsymbol{r} - \boldsymbol{r}_{ij})$ is Dirac's delta function, $f_c(\boldsymbol{r})$ is radial cutoff function. $\rho(\boldsymbol{r})$ is then expanded in the spherical coordinates:

$$c_{nlm} = \sum_{i \neq j} w(t_j)\, R_n(\boldsymbol{r}_{ij}) Y_{lm}(\theta, \phi) f_c(\boldsymbol{r}_{ij}) \tag{3}$$

The radial expansion function: $R_n(\boldsymbol{r}_{ij})$ takes the form of the linear transformed first-kind Chebyshev functions:

$$R_n(r_{ij}) = T_n\left(1 - \frac{2r_{ij}}{r_c}\right) \tag{4}$$

The linear transform is constructed so that the $R_n(r_{ij})$ is orthogonal in the range $[0, R_c]$, and the radial function increases monolithically as $r_{ij} \to 0$. While the expanding coefficients $C_{nlm}$ are complex numbers and not rotational invariant, the real part of the power spectrum:

$$P_{nl} = \frac{4\pi}{2l+1} \sum_{|m| \le l} c_{nlm}^{*} c_{nlm} \tag{5}$$

is rotationally invariant, which is suitable for use as structure descriptors.

For the efficient evaluating of $c_{nlm}$, the symmetry relation:

$$Y_{lm}(\theta, \phi) = (-1)^m Y_{l-m}^{*}(\theta, \phi) \tag{6}$$

is employed, which nearly halves the time of evaluation of the per-atom descriptor vectors and their derivatives.

While **Eq. 2-5** define structure descriptors for the multi-component systems if an appropriate weight function $w(t_j)$ is adequately selected[32], the selection of proper $w(t_j)$ values requires additional parameter-tuning work. Here we follow the idea from Artrith *et. al.*[28] and define a dual-basis representation for multi-component MGs. For the dual-basis representation, we first calculate the unweighted power spectrum: $P_{nl}^{s}$, where $w(t_j) \equiv 1$, then the weighted power spectrum $P_{nl}^{c}$ is calculated by using a simple weight function:

$$w(t_j) = (-1)^{t_j - 1} t_j, \qquad t_j = 1, 2, \cdots, n_t \tag{7}$$

where $n_t$ is the number of chemical species in the system. It is easy to conclude that the descriptor size is independent of $n_t$ for $n_t \ge 2$. We found that the dual-basis representation is more beneficial than manually tuning $w(t_j)$ values since the actual weight function can be "learned" automatically due to additional weight parameters in the neural networks.

For truncating descriptor values and derivatives to zero at the spherical cutoff $r_c$, a cutoff function $f_c(r)$ is needed. Currently, two kinds of cutoff functions are implemented in the NNAP code. The first one is:

$$f_c(r) = \begin{cases} \exp\left(\dfrac{\gamma (r/r_c)^2}{(r/r_c)^2 - 1}\right), & r < r_c \\ 0, & r \ge r_c \end{cases} \tag{8}$$

where $\gamma$ is a positive adjustable parameter to fine-tune the shape of the cutoff function. The second one is a simple parameter-free polynomial cutoff function:

$$f_c(r) = \begin{cases} (1 - (r/r_c)^2)^4, & r < r_c \\ 0, & r \ge r_c \end{cases} \tag{9}$$

It is simple to demonstrate that the zero to third derivatives are all smooth functions and take the values of zero at $r_c$ for both cutoff functions. Thus, both cutoff functions ensure smooth changes in atomic forces at cutoff when atoms enter or exit the cutoff radius, which is essential for reserving the energy stability in MD simulations[27].

The basic workflow for NNP construction procedure for MGs is organized as bootstrapping and refining stages, which is designed to improving the sample diversity of the final training dataset and minimizing the time-consuming AIMD samplings. An illustration of the overall workflow is presented in **Fig. S1** (see in the SI text).

In the bootstrapping stage, we first construct a preliminary database containing perturbated crystal phases and liquid configurations from AIMD simulations. The CUR algorithm[33] was used to select a subset of AIMD trajectory that the correlation of configurations is minimized. Then, the preliminary database is evaluated by the density functional theory (DFT) calculations, and the *k*-fold cross-validation is performed to obtain the initial *k*-NNPs ensemble. After that, the *k*-NNPs ensemble is iteratively optimized with an active-learning strategy as described in the supporting information.

In the refining stage, we sample the potential energy landscape (PEL) of the MGs to generate additional training samples. The sampling is done with different simulation protocols that generate samples from different parts of PEL. Currently, five protocols are used that containing: random structure search (RSS), MD, SHMC, basin hopping (BH), and activation relaxation technique *nouveau* (ARTn). The RSS is used to enhance sampling in ordered crystal phase space, which has been previously used for the MLP training[34]. The MD quenching is used to sample instantaneous and inherent configurations with a quenching rate larger than $10^{10}\ K/s$. The SHMC is used to obtain the deeply supercooled liquid and quenched MG samples with the effective quenching rate close to the laboratory-made MGs. A detailed description of SHMC simulations will be given in the following text. BH runs are mainly used for MG clusters which sample PEL minima around the global minimum[35]. ARTn runs provide the saddle point samples of MGs, which is essential for understanding the relaxations in MGs[36]. In practical application, ARTn is adopted only when the PEL is well represented by the current NNP so that the NNP-predicted saddle points are close to the true ones. For each part of the refining process, we iteratively perform the “training and DFT validation” cycle until the prediction error is comparable with the training error.

We select the ternary Zr-Cu-Al system as a representative of multi-component bulk metallic glasses to be used for illustrating our methods, which have been thoroughly studied over the past decades. **Figure S2a** shows the proportion of different data generation methods in the DFT training database. The part labeled “initial” corresponds to the bootstrapping stage in our workflow, and the other parts correspond to different refining methods in the refining stage. **Figure S2b** shows the distribution of training data in the Zr-Cu-Al ternary chemical space. The generated DFT training database contains 36,187 configurations corresponding to 1,794,343 atom environments, which almost covers the whole chemical space (**Fig. S2b**).

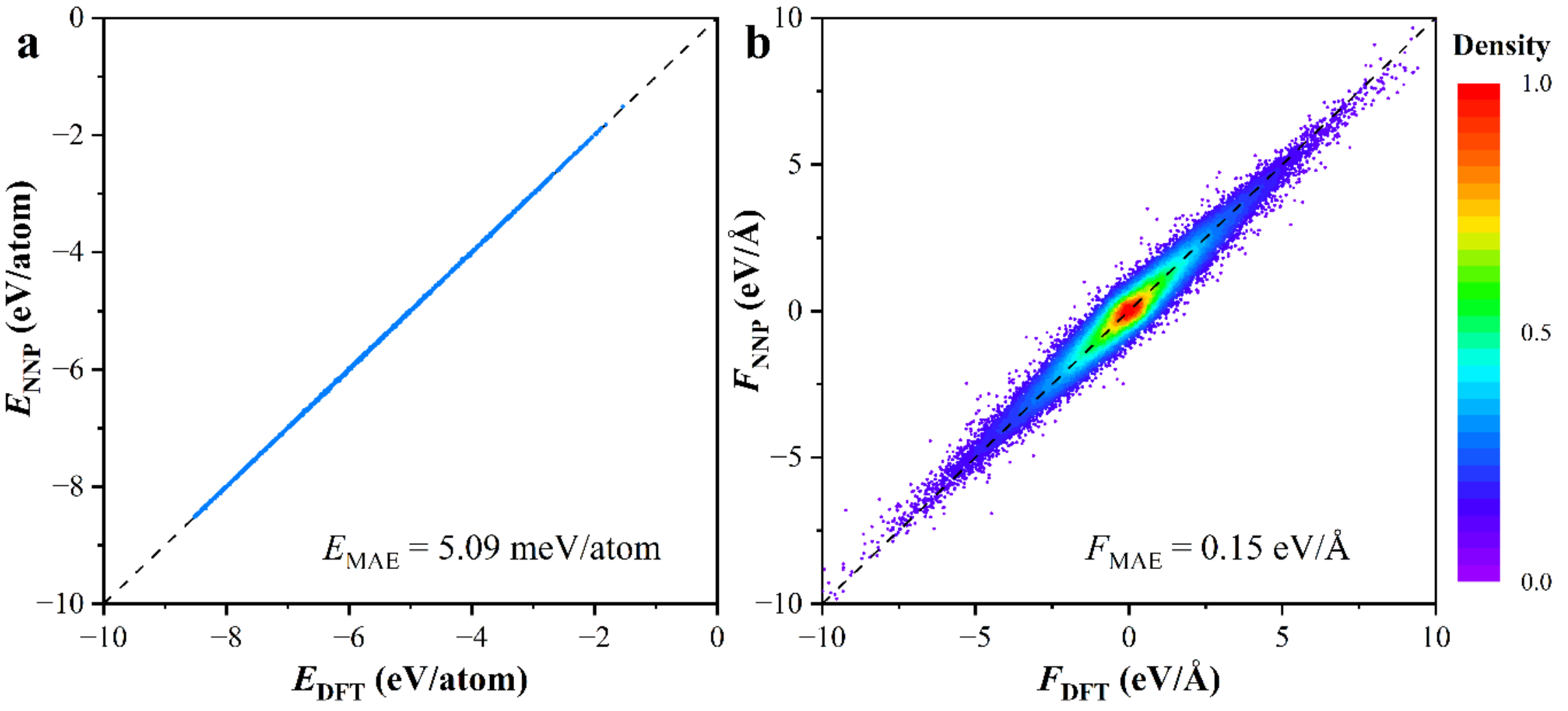

**Figure 1.** Calculated energies (a) and force components (b) fitting errors on the testing set. The perfect fitting: $y = x$ is shown as dashed lines. Data points in (b) are colored by their local number densities.

**Figure 1** shows the prediction results of the Zr-Cu-Al NNP on the testing set. The NNP achieves energy and force mean absolute errors (MAE) of 5.09 meV/atom and 0.15 eV/Å on the testing test, respectively. In comparison, the energy and force MAE on the training set is 4.42 meV/atom and 0.149 eV/Å in respective, which shows no overfitting.

Additionally, we carried out performance and transferability analysis (see in the SI texts) to show the computational efficiency and transferability of our NNP formalism and parameterization in respective.

### Atom Swapping assisted Hybrid Monte-Carlo (SHMC) simulation

One of the major drawbacks of NNP-based simulations is its highly complicated math form that hinders efficient energies and forces evaluations. Our performance analysis shows that the NNP-based MD simulations is about two orders of magnitudes slower than the EAM-based ones (see in the SI texts for more details). Thus, some kind of acceleration method is needed to apply NNP for studying the vitrification process as it is rather time-consuming even for the EAM potentials. In the past years, swapping Monte Carlo simulations have shown great ability to reach equilibrium in deeply supercooled liquids for poly-dispersed model glass-forming systems[22]. Given that most bulk MGs systems contain three or more components, it is straightforward to think that the same method could also apply to the NNP-based simulations of realistic MG samples. However, the direct application of swapping MC for MGs faces significant technical obstacles. Our practices show that most of the obstacles are come from the expensive NNP evaluations of energy and forces, which are not very important in the simple pair potentials used by model glass systems[23]. Swapping MC sweep consists of standard MC and random atom swapping MC. In the standard MC sweep, atoms are randomly displaced sequentially, which hinders efficient parallelization. For NNP simulations, the computation of the energy change due to single atom MC move requires a re-evaluation of $N$ times atom structure descriptor and the NN forwarding, where $N \approx 50 \text{ to } 100$ is close to the average number of per-atom neighbors. Since $3 \times N_{atoms}$ moves are required in a single MC sweep, it is impractical to simulate MG systems with more than several hundreds of atoms. To enhance the parallelization efficiency, the hybrid MD/MC method that combines conventional MD with atom-swapping MC has been suggested[37]. However, to ensure that the resulting sampling still follows the canonical probability distribution, the number of MD steps between two successive swapping MC steps should be larger than the timescale of thermostat. This typically means that more than 100 MD steps is required to ensure proper statistical distributions. As each MD step requires a full evaluation of potential energies and forces, such requirement enforces a strong performance limitation on NNP-based hybrid MD/MC simulations.

In this work, we propose to combine the hybrid Monte Carlo (HMC) method[38,39] with atom swapping, which we call SHMC method. The benefits of SHMC method are as follows: firstly, HMC make global updates of atom positions by taking short-time MD steps that can be efficiently parallelized; secondly, the MD parameters of HMC can be arbitrary selected without affecting the equilibration distribution. The HMC sweeps make global updates of atom positions by taking short-time MD steps with initial atom velocities randomly drawn from a Gaussian distribution at a given temperature $T$. The MC update is accepted with a transition probability[39]:

$$P_{acc}(\boldsymbol{x};\boldsymbol{v} \to \boldsymbol{x}';\boldsymbol{v}') = \min\left\{1, e^{-\beta\Delta H}\right\} \tag{10}$$

where $\boldsymbol{x};\boldsymbol{v}$ and $\boldsymbol{x}';\boldsymbol{v}'$ are successive points in the phase space, $\beta = 1/k_{\mathrm{B}}T$ is the Boltzmann factor, and $\Delta H$ is the change of Hamiltonian (total energy) due to the MD update. The *detailed balance* of such MD update is ensued if the numerical integration of the motion equations is *time-reversible* and *area-preserving*[39]. Here we use the *leap-frog* method for integrating the motion equations. Since the *detailed balance* is obeyed, the whole simulation can be viewed as a Markov process so that the canonical probability distribution of the system is solely determined by **Eq. 10**, which does not depend on the MD parameters, especially the number of MD steps.

Unlike model glass systems, simulations of MGs are usually performed under the *isothermal isobaric* (*NpT*) ensemble. While generalizing current MD schemes to the *NpT* ensemble is possible[39], we use additional volume-scaling MC steps[40] to maintain the system pressure:

$$P_{acc}(V_0 \to V) = \min\{1, \exp[-\beta(\delta U + P(V - V_0) - N\beta^{-1}ln(V/V_0))]\} \tag{11}$$

The atom-swapping MC steps are performed by exchanging the chemical symbols of a pair of randomly chosen atoms. Here, there are two methods to choose the atom pairs to be swapped. The first method (method I) involves choosing an atom A at random, followed by choosing an atom B with a different chemical symbol from atom A at random. In practical, the method I is supplied by the “early rejection” so that the atom exchange with a large size difference is rejected without energy evaluation. However, since there are only three elements in our system, directly reject Cu-Zr exchange would miss events that contribute to large structure relaxations. Thus, we propose the second method (method II) that involves randomly choosing elements, A and B. from the composition under investigation, followed by a random selection of an atom from each element A and B in the configuration. By lowering the ratio of Cu-Zr exchange, the method II could achieve a higher acceptance ratio in practical MGs simulation, and it is also insensitive to the chemical composition of MGs. For example, considering the 100-atom MG system $Zr_{46}Cu_{46}Al_8$, the probability of the Cu-Al exchange event is calculated as $((46/100) \times (8/54) + \big((8/100) \times (46/92)\big) \approx 0.11$ for the method I that depends on the detailed chemical composition, but the probability of the same event for method II is $2 \times (1/3) \times (1/2) \approx 0.33$ that is a constant value that is independent of the exact chemical composition.

As the potential energy changes due to atom swapping are evaluated many times in the SHMC simulations, the evaluation of potential energy difference must be aggressively optimized so that SHMC simulations can be efficiently performed. For example, a $10^3$-atom SHMC run of $10^6$ sweeps with 300 swapping attempts in each sweep would require $3 \times 10^8$ times evaluations of potential energy differences. A full evaluation of the total potential energy before and after each swapping attempt is computationally unacceptable as they scale linearly with system size. However, it is often unnecessary. For the worst-case estimation, if we assume the maximum number of neighbors is $N_{neigh}$ and the neighbors of atoms A and B do not overlap, only parts of the per-atom descriptors would change due to A-B swapping and the maximum number of required per-atom energy evaluations $N_{eval} = \min\{N_{atoms}, 2 \times (N_{neigh} + 1)\}$. If the system is large enough, the computation complexity of each atom swap attempt is not dependent on the system size $N_{atoms}$. To further reduce the computation cost, the per-atom local potential energies are calculated and saved in memory before the swapping part of the MC sweep. During each swapping attempt, the local potential energies are updated in case of successful swapping attempts. This optimization saves one energy evaluation for each swap attempt and nearly halves the number of required potential energy evaluations in the atom swapping MC. Additionally, the energy evaluation is fully parallelized with an atom-decomposition method.

In our NNP-SHMC code, one MC sweep of the SHMC simulation consists of three sequential parts: short MD run, volume scaling (only for the *NpT* runs) and atom swapping. The number of atom-swapping attempts is determined by a single parameter $r_{swap}$ as $N_{swap} = r_{swap} \times N_{atoms}$. Thus, SHMC simulations require three main parameters: the number of MD steps ($N_{MD}$), the timestep size of MD run ($t_{MD}$) and the number of swap attempts ($r_{swap}$) at each MC sweep. As we have discussed in the above text, the target canonical probability distribution at a given temperature is not altered by these parameters as the *detailed balance* condition is always satisfied in the simulation, but the structural relaxation time would be strongly affected by the combinations of the three parameters and should be fine-tuned for the maximum acceleration of SHMC runs.

### Thermodynamical stabilities of Zr-Cu crystalline phases

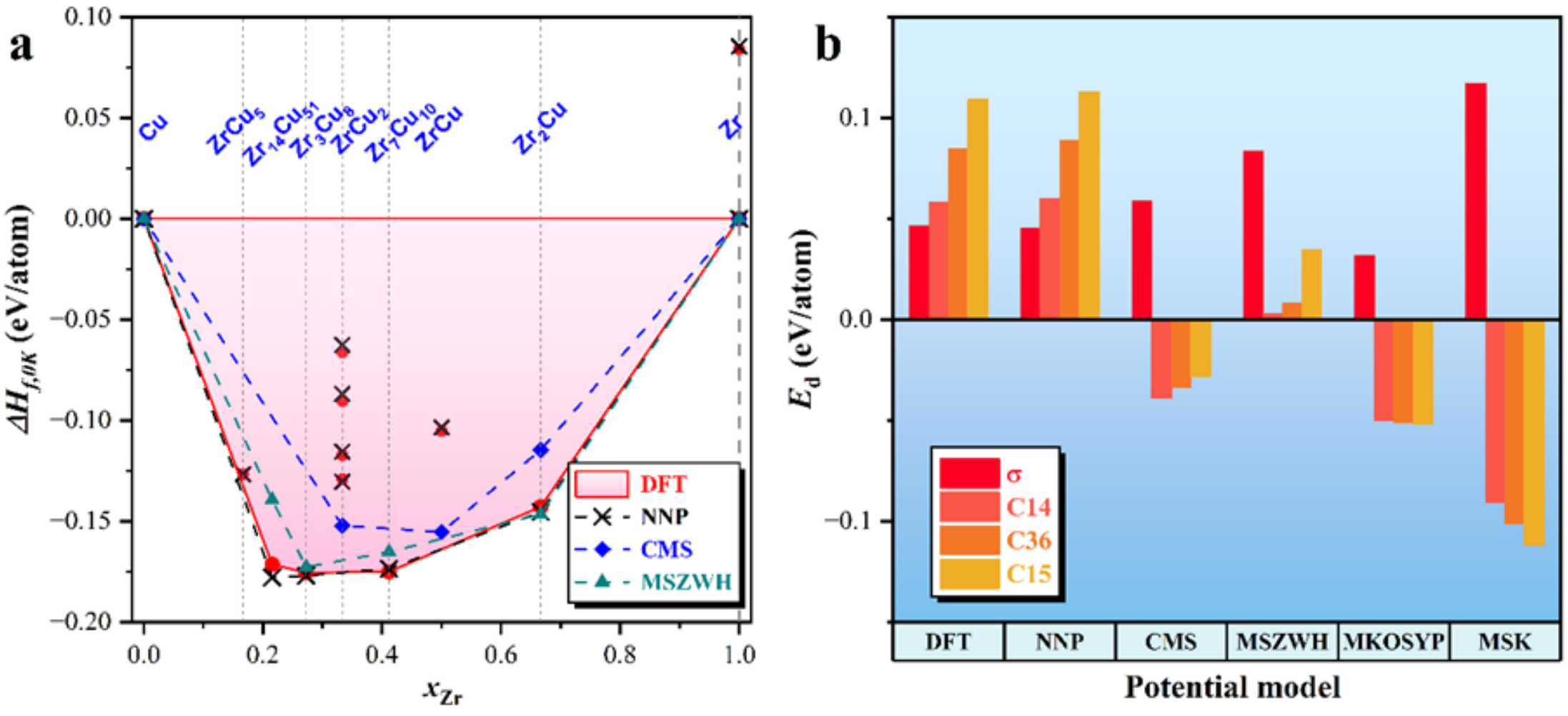

**Figure 2.** Zr-Cu composition diagrams (**a**) and decomposition energies $E_d$ (**b**) of $Zr_2Cu$ phases obtained from various potentials. For the CMS and MSZWH potentials, only stable phase boundaries are shown in (**a**). The $ZrCu_2$ decomposition energy is defined as: $E_d \equiv (13E_{ZrCu_2} - 2E_{Zr_3Cu_8} - E_{Zr_7Cu_{10}})/39$.

We compare the Zr-Cu composition diagrams obtained from different potential models in **Fig. 2** to validate the NNP performances on the crystalline phases. The full results for all investigated crystalline phases, including also Zr-Al and Cu-Al phases, are listed in **Table S1**. The NNP results are compared with the ones from widely used Zr-Cu-(Al) EAM potentials: MSK[41], MKOSYP[42] , MSZWH[18] and CMS[43] (abbreviated by their authors' names in respective). Among all the investigated models, our NNP model gives almost identical predictions of Zr-Cu phase energies as DFT, which has an MAE of ~2 meV/atom (see **Table S1**). Surprisingly, none of the previous EAM models can fully predict the correct Zr-Cu phase stabilities. The CMS potential, as well as MSK and MKOSYP ones (not shown), is well known[18,44] to incorrectly predict the ZrCu and $ZrCu_2$ phases as stable ones. While the recently developed MSZWH potential gives correct phase boundaries, the derivation to DFT results is very large for the Cu-rich $Zr_{14}Cu_{51}$ phase, which owns a complex crystal structure that is hard to be described by the simplified EAM model. Importantly, none but the NNP model can produce the correct energy sequence of $ZrCu_2$ metastable phases (**Fig. 2b**).

The $ZrCu_2$ composition is a well-known glass-forming liquid[45] in the Zr-Cu phase diagram. The energy sequence of metastable phases is crucial to understand the vitrification and crystallization processes in Zr-Cu liquids. As shown in **Fig. 2b**, all EAM models predict some (C14 or C15) of the Laves phases as the most stable $ZrCu_2$ phase, while DFT and NNP show that the $\sigma$ phase is the most stable one. From the view of local structures, the Laves phases (C14, C36, and C15) all correspond to Cu-centered icosahedrons, while the $\sigma$ phase corresponds to a distorted bcc-like structure that does not contain any icosahedrons. Thus, the EAM potentials might overestimate the stability of such icosahedrons in simulations. The detailed influences of such differences on the GFA and crystallization pathway of $ZrCu_2$ are worthy of further study and will be presented in future works.

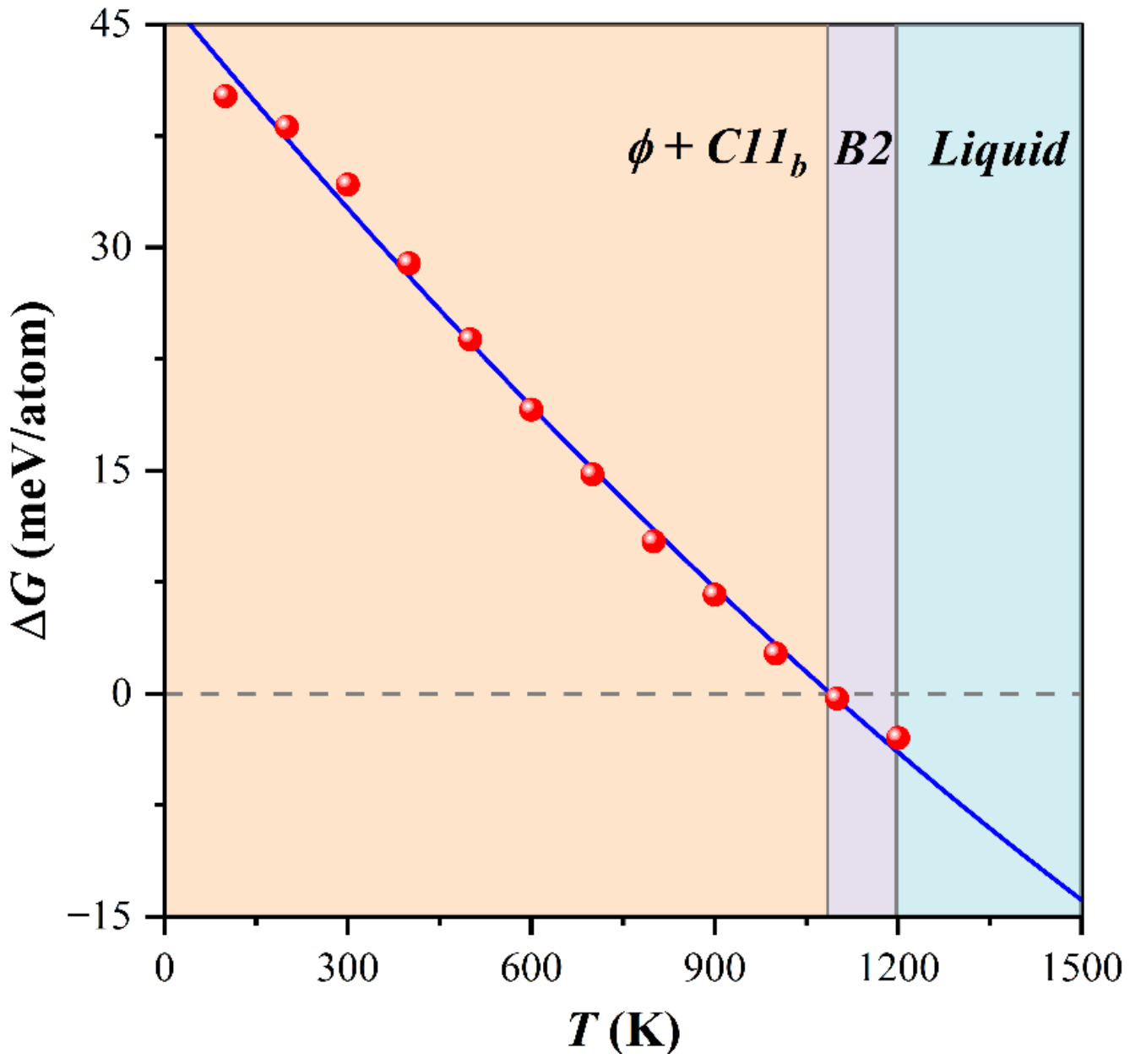

**Figure 3.** Free energy difference plot showing the thermodynamically stable region of B2 phase. The red dots show the calculated points, and the blue line indicates the quadratic fitting.

**Table 1.** Phase decomposition temperature $T_\mathrm{d}$ and melting temperature $T_\mathrm{m}$ for B2-ZrCu.

| | Exp. | NNP | CMS | MSZWH | MKOSYP | MSK |
|---|---|---|---|---|---|---|
| $T_\mathrm{d}\ (K)$ | 985-1003 | 1090 | - | 190 | - | - |
| $T_\mathrm{m}\ (K)$ | 1209-1229 | 1197 | 1347 | 989 | 1353 | 1741 |

The B2-ZrCu phase corresponds to another glass-forming composition[45] on the phase diagram and is the parent structure for Zr-Cu-Al MGs. The B2-ZrCu has been thoroughly investigated for the thermodynamical evolution during the crystallization[46,47]. Previous modeling[48] has shown that the melting temperature of B2 phase was significantly underestimated by some EAM models. In **Fig. 3**, we presented the NNP-predicted free energy curve and melting temperature, which determine the thermodynamically stable region of the B2-ZrCu phase. The NNP, EAM, and experimental results are summarized in **Table 1**. In this work, the free energies are computed by the Frenkel-Ladd method[49] at various temperatures, and the melting temperatures are calculated through the "interface pinning" method[50,51]. The details on the free energy and melting temperature calculations are presented in the supplemental materials. Of all the examined potentials, only NNP and MSZWH can describe the meta-stability of the B2 phase. However, the NNP reproductions are in good agreement with the experiments, whereas the actual numbers of the MSZWH reproductions deviate significantly from the experimental measurements[46]. As a result, the NNP would enable accurate reproductions of the Zr-Cu phase stabilities that are not possible with nowadays EAM potentials.

**Structural properties of liquid and amorphous Zr-Cu-Al configurations**

The pair correlation functions $g(r)$ and structure factors $s(q)$ for the $Zr_{54}Cu_{46}$ and $Zr_{46}Cu_{46}Al_8$ systems are calculated to validate the NNP performance on the structural properties at liquid and amorphous states. **Figure 4a** shows the calculated $g(r)$ compared with the DFT-MD results. The DFT results are obtained through 100-atom configurations with time durations of $15\,ps$, while the NNP results are obtained through 1,000-atom configurations with time durations of $200\,ps$. The NNP results are in good agreement with the DFT ones for both $Zr_{54}Cu_{46}$ and $Zr_{46}Cu_{46}Al_8$. Experimentally, $s(q)$, rather than $g(r)$, are directly obtained from scattering experiments. We compare calculated $s(q)$ curves to the experimentally measured results[52,53] in **Fig. 4b**. For the direct comparison to the laboratory-made $Zr_{46}Cu_{46}Al_8$ amorphous sample, we use the SHMC-generated well-annealed MG models for calculating the room temperature $s(q)$. More details on these models will be presented in the following text. In both compositions, the results compare well to the experimental measurements, while slightly underestimated first-peak strength can be observed. The minor differences might come from the sample-size and energy-stability discrepancies between experiments and simulations. More important, the splitting of the second $s(q)$ peak in $Zr_{46}Cu_{46}Al_8$ is well reproduced by the NNP sample. It has been shown that the splitting of the second peak corresponds to the development of medium-range structural order in MGs[54], which hints that our NNP also provides a good description of the medium-range order of Zr-Cu-Al MGs. The detail about the accurate atomic packings is another important research topic to be investigated in the future.

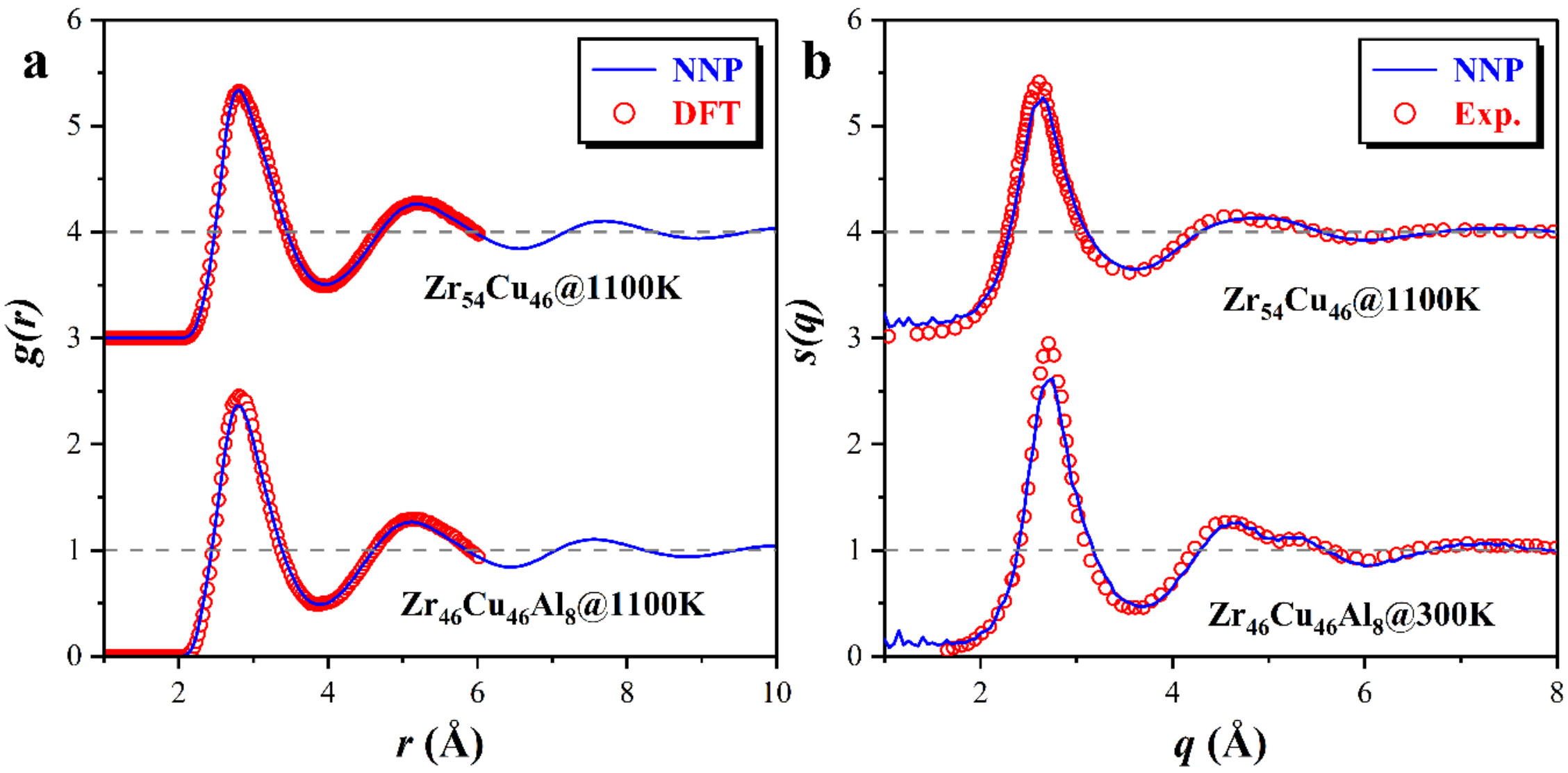

**Figure 4.** Structural properties for liquid and amorphous Zr-Cu-Al configurations. (**a**): Pair correlation functions $g(r)$ calculated by NNP (solid blue lines) and DFT (open red circles), respectively. (**b**): Static structural factor $s(q)$ obtained by NNP-MD or NNP-SHMC simulations (solid blue lines) and experimental measurements (open red circles), respectively.

### SHMC-Accelerated relaxation dynamics in the deeply supercooled Zr-Cu-Al liquids

To prepare the atomic configurations of MGs with an effective cooling rate close to the laboratory one, we need to achieve the thermodynamic equilibrium of the deeply supercooled liquids in the simulations. However, the viscosity of deeply supercooled liquids increases dramatically towards the glass transition temperature $T_g$, making it nearly impossible to achieve equilibrium in deeply supercooled liquids through standard MD simulations. As discussed in above text, we propose the SHMC method that is expected to accelerate the relaxation dynamics and realizing the thermodynamic equilibrium process of deeply supercooled liquid. Here, we systematically investigate the temperature-dependent relaxation dynamics of $Zr_{46}Cu_{46}Al_8$ glass-forming liquid by using the NNP-SHMC method. To characterize the structural relaxations in liquids, the $\alpha$-relaxation time $\tau_\alpha$ is extracted from the self-intermediate-scattering function (SISF) $F_s(\boldsymbol{q},t)$:

$$F_s(\boldsymbol{q},t) = \left\langle \frac{1}{N}\sum_j e^{i\boldsymbol{q}\cdot[\boldsymbol{r}_j(t)-\boldsymbol{r}_j(0)]} \right\rangle \quad (12)$$

where $\boldsymbol{q} = 2.67\ \text{Å}^{-1}$ is used, and the $\tau_\alpha$ is defined as $F_s(q,\tau_\alpha) = e^{-1}$.

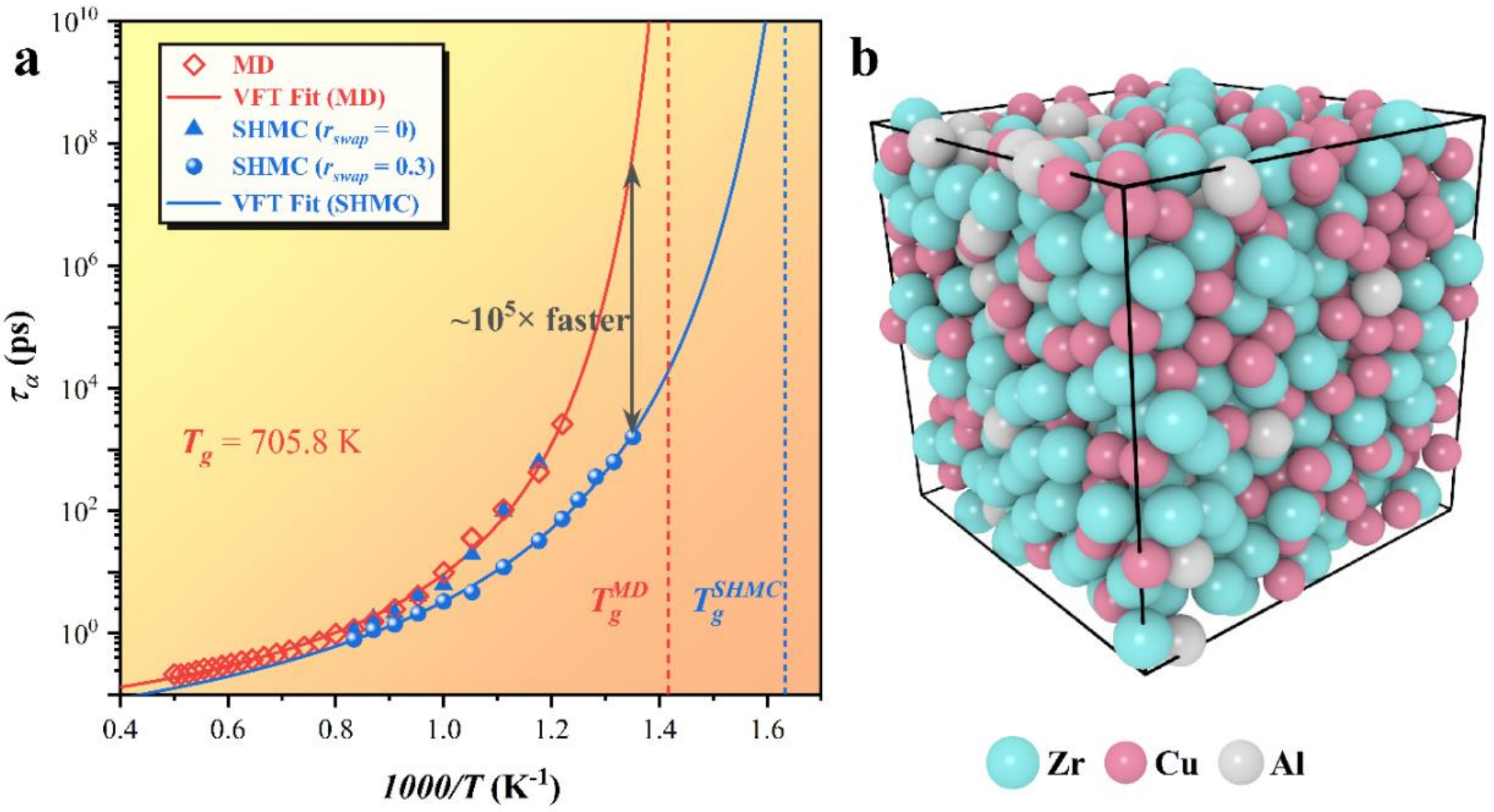

**Figure 5.** (**a**): Extracted structural relaxation time $\tau_\alpha$ with inverse temperature $1000/T$ for standard MD simulations (red open squares) and SHMC ones with $r_{swap} = 0$ (solid red triangles) and $0.3$ (blue balls), respectively. An effective time scale $t_0 = 25\ fs$ is used to scale the SHMC runs. Vogel–Fulcher–Tammann (VFT) fittings for MD (solid red line) and SHMC (solid blue line) runs are presented to estimate the acceleration ratios. (**b**): Snapshot of SHMC-equilibrated sample at $T = 740\ K$.

As shown in **Fig. 5a,** we analyze the structural relaxation time $\tau_\alpha$ evolutions with inverse temperatures under different simulation protocols using the Zr-Cu-Al NNP. Extrapolation of the Vogel–Fulcher–Tammann (VFT) fitting on the MD results to a typical time scale: $\tau_\alpha = 100\ s$ predicts the glass-transition temperature ($T_g$) of $Zr_{46}Cu_{46}Al_8$ as $705.8\ K$, which is in excellent agreement with experiment result that $T_g^{\text{exp}} \approx 709\ K$[53]. This further indicates that our NNP potential can accurately reproduce the thermodynamic properties of the Zr-Cu-Al supercooled liquids. As SHMC simulations are performed through MC sweeps, an effective time scale must be estimated for direct comparison with the MD results since time is not well-defined in SHMC runs. As shown in **Fig. 5a**, the nice scaling collapse between the MD and swapping-free ($r_{swap} = 0$) SHMC results allows us to evaluate the effective time scale $t_0 = 25\ ps$ for each MC sweep. As atom swapping is introduced, the equilibration process of metallic glass-forming liquids is obviously accelerated in the deeply supercooled liquid region. The acceleration ratio increases dramatically with the decreasing of liquid temperatures. At the lowest investigated temperature $T = 740\ K$, the acceleration ratio reaches a value of $10^5$ times. Consider that the NNP-SHMC run at $740\ K$ with $2 \times 10^6$ MC sweeps using 108 CPU cores takes about five days, the MD run would take about 1,370 years to reach the same equilibrium, which is far out of current computing powers. The significant acceleration of the relaxation dynamics of deeply supercooled liquids easily fills the performance gap between EAM and NNP potentials, which enables us to efficiently sample the phase space that accelerates the vitrification process of the multi-component metallic glass-forming liquid to a time scale close to the laboratory conditions.

**Vitrification of Zr-Cu-Al MG from deeply supercooled liquids**

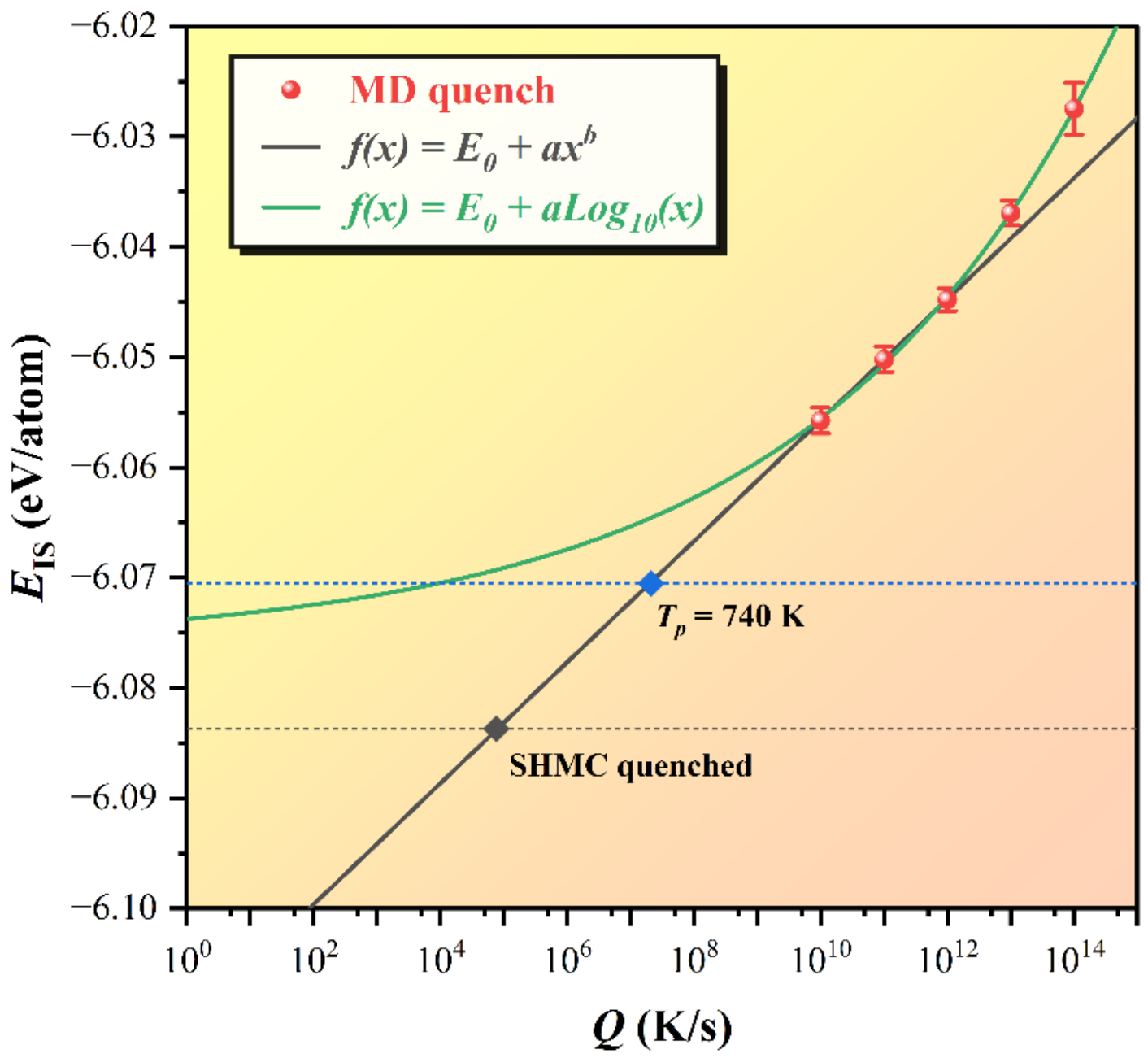

**Figure 6.** Estimating effective quenching rates for SHMC-prepared samples. Red balls present the inherent structure energies $E_{\mathrm{IS}}$ for MD-prepared samples from five quenching rates: $10^{14}, 10^{13}, 10^{12}, 10^{11}$, and $10^{10}\ K/s$. Error bars are calculated from 5-10 independent MD runs. Solid black line shows the logarithmic extrapolation of the lower part of MD data. Solid green line shows the power-law extrapolation. Extrapolated points for the equilibrated sample at $T_P = 740\ K$ and the SHMC quenched one are shown as filled blue and black diamonds, respectively.

When the deeply supercooled $Zr_{46}Cu_{46}Al_8$ liquid reaches its thermodynamic equilibrium at 740K, no crystallization is observed, as shown in **Fig. 5b**. This fact indicates that the NNP provides the accurate description of the glass-forming ability of the MG system, which allows the out-of-equilibrium vitrification from the equilibrium liquids to be simultaneously accelerated by the NNP-SHMC method. Thus, it is interesting to check the stability of the simulated MG samples due to the NNP-SHMC accelerated vitrification process. To this end, we perform NNP-SHMC-based vitrification simulation from the equilibrium liquid at $T = 740\ K$.

The convention of the glass physics community uses the “parent temperature” $T_P$, where the glass samples are prepared by minimization from the supercooled liquid, to characterize the glass stability of samples[22,55,56]. In contrast, the quenching rate $Q$ is usually used to characterize the thermal stability of MG samples experimentally, as it is more convenient to be measured. Here we estimate the effective quenching rate of the SHMC-prepared MG sample by comparing the inherent structure energy $E_{\mathrm{IS}}$ of the SHMC-prepared MG samples with the MD-prepared ones. The inherent structure energy $E_{\mathrm{IS}}$ has been previously used to define the absolute "glass stability" or "annealing degree"[57] that also well correlates with the parent temperature $T_P$[55]. The MD and SHMC results are presented in **Fig. 6**. Since the analytic relationship between $E_{\mathrm{IS}}$ and $Q$ is unknown, extrapolations using two functional forms are performed to estimate the effective $Q$ values. One is the simplest logarithmic extrapolation. As the MD results at very high ($> 10^{12}\ K/s$) quenching rates break the logarithmic relationship, only the last three MD points in **Fig. 6** are used for the extrapolation. Alternatively, the power-law relationship proposed in reference[58] is applied to use all MD data points.

In **Fig. 6**, $E_{\mathrm{IS}}$ values of two metallic glasses samples are presented. One corresponds to the sample equilibrated at "parent temperature" $T_p = 740\ K$, denoted as the 740K-equilibrated one. The other corresponds to the SHMC-accelerated vitrification from 740 K to 300 K in $1 \times 10^6$ MC sweeps under a constant external pressure $P = 0\ GPa$, denoted as the SHMC-quenched one. For the 740K-equilibrated sample, logarithmic extrapolation yields an effective $Q_e \approx 10^7\ K/s$, while the power-law extrapolation yields $Q_e \approx 10^4\ K/s$. However, power-law extrapolation cannot estimate $Q_e$ for the SHMC-quenched sample since the fitted power-law function is always larger than the sample's $E_{\mathrm{IS}}$. The departure of the two extrapolations comes from the lack of MD reference data at a lower quenching rate that cannot be obtained due to the wall time limit for nowadays computers. Instead, the $Q_e$ of the SHMC-quenched sample is estimated to be $\sim 1 \times 10^5\ K/s$ by the logarithmic extrapolation, which is already less than the maximum experimental quenching rate ($\sim 10^6\ K/s$) of the melt-spinning MG ribbons[59]. Moreover, the calculated structural factor $s(q)$ of the sample (as shown in **Fig. 4b**) still exhibits the characteristics of the amorphous structure and agrees well with the neutron diffraction experiment results. Thus, the developed NNP-SHMC protocol enables us to obtain MG samples and atomic-level properties of stability comparable with the laboratory-made multi-component MG samples.

# Discussion

Accurate atomic structure model of metallic glass-forming liquid at deeply supercooled temperatures is the key to understanding the structural origin of the dynamical arresting[6,11]. Due to the limitation of DFT calculations, EAM-based potentials are developed that contribute most of our current knowledge of the vitrification process. Developing accurate and transferable EAM potentials require hard work to optimizing the model parameters and is error-prone in the selection of the reference datasets[18]. In this work, we suggest and detailly analyze the novel NNP-SHMC method to generate accurate and well-annealed MG samples. With a simple and computationally efficient descriptor form, we achieve accurate and efficient NNP modeling of the representative Zr-Cu-Al bulk MGs with a diverse DFT training database that expands extended energy and composition space. The accuracy of NNP is illustrated by comparing many thermodynamical and structural properties of Zr-Cu-Al crystalline and glass phases against experimental results. Additionally, performance and transferability analysis (see more in SI texts) show that the current NNP formalism is computational efficient and transferable. With the powerful SHMC simulation, we show that the thermodynamic equilibrium of a multi-component metallic glass-forming liquid can be achieved at the deeply supercooled temperature close to $\boldsymbol{T_g}$ within several days, which might need thousands of years for conventional MD simulations. The SHMC-quenched MG sample shows an effective quenching rate comparable with the laboratory-made MG samples, which allows direct comparison of the NNP-generated samples with future experiments.

In conclusion, our findings pave the way for understanding the composition-dependent physical properties, such as atomic structure, glass forming ability, and mechanical behaviors of laboratory-produced metallic glasses, and thereby the computational design of compositionally complex alloys. For example, the accurate inter-atomic interactions and large-sized samples with high thermal stability enable better capture of the effects due to small changes in chemical composition, which makes it possible to study the origin of trace-element effects on the macroscopic properties in amorphous alloys with the NNP-SHMC method. Furtherly, extending NNP-SHMC to other multi-component MGs or high entropy alloys (HEA) is straightforward and should help to clarify the atomic-level solidification processes of MGs or HEAs. Moreover, as NNP-SHMC generates samples close to the DFT reference, it can provide atomic configurations for the electronic structure calculations, thus contributing to the understanding and establishment of the intrinsic correlations between local electronic properties and local chemical environment.

# Materials and Methods

### NNP training

The SC basis is used as the structural descriptor. All three elements share the same parameters for the SC basis: $n_{\max} = 5$, $l_{\max} = 6$ and $r_{\mathrm{c}} = 6.5$ Å, and the polynomial cutoff function (Eq. 9) is selected. For each specie, we use a three-hidden-layer NN architecture of $84 \times 16 \times 16 \times 16 \times 1$, which results in 5,763 degrees of freedom in total. The activation function is set to the sigmoid linear unit (SiLU) function: $\mathrm{SiLU}(x) = x/(1 + e^{-x})$. All training is performed by the LBFGS-B[60] optimizer implemented in our in-house NNAP code and a $\ell_2$-regulation factor $\lambda = 0.002$ is used to reduce overfitting. A 10-fold cross-validation run is performed to generate the NNP ensembles. The DFT potential energies are used in NNP training.

### DFT calculations

All DFT calculations are performed by the Vienna *ab initio* simulation package[61] (VASP). The electron-nuclear interaction is described by the all-electron projector augmented wave[62] (PAW) method. The PBE[63] exchange-correlation functional is used to describe the electron-electron correlation energies under the generalized gradient approximation (GGA). A plane wave cutoff of 420 eV is used for expanding the wave functions. A uniform K-point mesh with a spacing of $0.3$ Å$^{-1}$ is used for reciprocal space integrations of periodic structures, otherwise a single $\Gamma$ point is used. The DFT energy convergence criterion is set to $1 \times 10^{-5}$ eV.

**NNP-MD simulations**

NNP-MD simulations are performed by the LAMMPS code[64] with the in-house LAMMPS-NNAP interface. The MD time step is set to 2 fs, and a constant pressure and temperature (NpT) ensemble is employed in which the temperature and pressure were controlled by the Nosé-Hoover[65,66] and Parrinello-Rahman[67] methods, respectively.

**NNP-SHMC simulations**

NNP-SHMC simulations are performed by our in-house NNAP code. The three main parameters: $N_{MD}$, $t_{MD}$, and $r_{swap}$, are set to 8, 16.0 fs, and 0.3, in respective. The parameters are optimized for achieving maximum accelerations. More information about the optimization of parameters is available in the SI texts.

# Data availability

The full DFT training database of the Zr-Cu-Al systems is available from the corresponding author on reasonable request.

# Code availability

The NNAP code used in this study is under active development and not publicly available at this time but may be made available to qualified researchers on reasonable request from the corresponding author.

# Acknowledgments

This work is supported by the NSF of China (Grant Nos. 5211101002 and U1930402) and the National Key R&D Program of China (Grant No. 2017YFA0303400). R.S. acknowledges the Young Scientists Fund of the National Natural Science Foundation of China (No. 51801046). P.F.G. and R.S. acknowledge the computational support from the Beijing Computational Science Research Center (CSRC). Additional computational resources from the Institute of Advanced Magnetic Materials of Hangzhou Dianzi University are also gratefully acknowledged.

# Author contributions

R.S., and P.F.G. designed the research; R.S. wrote the NNAP code; R.S performed DFT calculation and constructed the Zr-Cu-Al potentials; R.S., Y.J.Y., P.F.G., and W.H.W. analyzed the results. The manuscript is written by R.S. and P.F.G. All authors contributed to the discussions and comments on the manuscript.

# Competing interests

All authors declare no financial or non-financial competing interests.

# References

1. Chen, M. Mechanical behavior of metallic glasses: microscopic understanding of strength and ductility. *Annu. Rev. Mater. Res.* **38**, 445–469 (2008).

2. Demetriou, M. D. *et al.* A damage-tolerant glass. *Nat. Mater.* **10**, 123–128 (2011).

3. Zhao, M., Abe, K., Yamaura, S., Yamamoto, Y. & Asao, N. Fabrication of Pd–Ni–P metallic glass nanoparticles and their application as highly durable catalysts in methanol electro-oxidation. *Chem. Mater.* **26**, 1056–1061 (2014).

4. Hu, Y. C. *et al.* A highly efficient and self-stabilizing metallic-glass catalyst for electrochemical hydrogen generation. *Adv. Mater.* **28**, 10293–10297 (2016).

5. Klement, W., Willens, R. H. & Duwez, P. Non-crystalline structure in solidified gold–silicon alloys. *Nature* **187**, 869–870 (1960).

6. Royall, C. P. & Williams, S. R. The role of local structure in dynamical arrest. *Phys. Rep.* **560**, 1–75 (2015).

7. Sheng, H. W., Luo, W. K., Alamgir, F. M., Bai, J. M. & Ma, E. Atomic packing and short-to-medium-range order in metallic glasses. *Nature* **439**, 419–425 (2006).

8. Hirata, A. *et al.* Geometric frustration of icosahedron in metallic glasses. *Science* **341**, 376–379 (2013).

9. Hirata, A. *et al.* Direct observation of local atomic order in a metallic glass. *Nat. Mater.* **10**, 28–33 (2011).

10. Guan, P. F., Fujita, T., Hirata, A., Liu, Y. H. & Chen, M. W. Structural origins of the excellent glass forming ability of $Pd_{40}Ni_{40}P_{20}$. *Phys. Rev. Lett.* **108**, 175501 (2012).

11. Hu, Y. C., Li, F. X., Li, M. Z., Bai, H. Y. & Wang, W. H. Five-fold symmetry as indicator of dynamic arrest in metallic glass-forming liquids. *Nat. Commun.* **6**, 8310 (2015).

12. Wang, B. *et al.* Understanding atomic-scale features of low temperature-relaxation dynamics in metallic glasses. *J. Phys. Chem. Lett.* **7**, 4945–4950 (2016).

13. Xu, B., Falk, M. L., Li, J. F. & Kong, L. T. Predicting shear transformation events in metallic glasses. *Phys. Rev. Lett.* **120**, 125503 (2018).

14. Hu, Y. C. *et al.* Configuration correlation governs slow dynamics of supercooled metallic liquids. *Proc. Natl. Acad. Sci.* **115**, 6375–6380 (2018).

15. Francis, G. P. & Payne, M. C. Finite basis set corrections to total energy pseudopotential

calculations. *J. Phys. Condens. Matter* **2**, 4395–4404 (1990).

16. Murali, P. *et al.* Atomic scale fluctuations govern brittle fracture and cavitation behavior in metallic glasses. *Phys. Rev. Lett.* **107**, 215501 (2011).
17. He, Y., Yi, P. & Falk, M. L. Critical analysis of an FeP empirical potential employed to study the fracture of metallic glasses. *Phys. Rev. Lett.* **122**, 035501 (2019).
18. Mendelev, M. I., Sun, Y., Zhang, F., Wang, C. Z. & Ho, K. M. Development of a semi-empirical potential suitable for molecular dynamics simulation of vitrification in Cu-Zr alloys. *J. Chem. Phys.* **151**, 214502 (2019).
19. Artrith, N. & Urban, A. An implementation of artificial neural-network potentials for atomistic materials simulations: Performance for TiO2. *Comput. Mater. Sci.* **114**, 135–150 (2016).
20. Zhang, L., Han, J., Wang, H., Car, R. & E, W. Deep potential molecular dynamics: a scalable model with the accuracy of quantum mechanics. *Phys. Rev. Lett.* **120**, 143001 (2018).
21. Wen, T. *et al.* Development of a deep machine learning interatomic potential for metalloid-containing Pd-Si compounds. *Phys. Rev. B* **100**, 174101 (2019).
22. Ninarello, A., Berthier, L. & Coslovich, D. Models and algorithms for the next generation of glass transition studies. *Phys. Rev. X* **7**, 021039 (2017).
23. Parmar, A. D. S., Ozawa, M. & Berthier, L. Ultrastable metallic glasses in silico. *Phys. Rev. Lett.* **125**, 085505 (2020).
24. Parmar, A. D. S., Guiselin, B. & Berthier, L. Stable glassy configurations of the Kob–Andersen model using swap Monte Carlo. *J. Chem. Phys.* **153**, 134505 (2020).
25. Behler, J. & Parrinello, M. Generalized neural-network representation of high-dimensional potential-energy surfaces. *Phys. Rev. Lett.* **98**, 146401 (2007).
26. Yanxon, H., Zagaceta, D., Wood, B. C. & Zhu, Q. Neural networks potential from the bispectrum component: A case study on crystalline silicon. *J. Chem. Phys.* **153**, 054118 (2020).
27. Behler, J. Atom-centered symmetry functions for constructing high-dimensional neural network potentials. *J. Chem. Phys.* **134**, 074106 (2011).
28. Artrith, N., Urban, A. & Ceder, G. Efficient and accurate machine-learning interpolation of atomic energies in compositions with many species. *Phys. Rev. B* **96**, 014112 (2017).
29. Gastegger, M., Schwiedrzik, L., Bittermann, M., Berzsenyi, F. & Marquetand, P. wACSF—Weighted atom-centered symmetry functions as descriptors in machine learning potentials. *J. Chem. Phys.* **148**, 241709 (2018).
30. Bartók, A. P., Kondor, R. & Csányi, G. On representing chemical environments. *Phys. Rev. B* **87**, 184115 (2013).
31. Bartók, A. P., Payne, M. C., Kondor, R. & Csányi, G. Gaussian approximation potentials: the

accuracy of quantum mechanics, without the electrons. *Phys. Rev. Lett.* **104**, 136403 (2010).

32. Willatt, M. J., Musil, F. & Ceriotti, M. Feature optimization for atomistic machine learning yields a data-driven construction of the periodic table of the elements. *Phys. Chem. Chem. Phys.* **20**, 29661–29668 (2018).

33. Imbalzano, G. *et al.* Automatic selection of atomic fingerprints and reference configurations for machine-learning potentials. *J. Chem. Phys.* **148**, 241730 (2018).

34. Deringer, V. L., Pickard, C. J. & Csányi, G. Data-driven learning of total and local energies in elemental boron. *Phys. Rev. Lett.* **120**, 156001 (2018).

35. Wales, D. J. & Doye, J. P. K. Global optimization by basin-hopping and the lowest energy structures of Lennard-Jones clusters containing up to 110 atoms. *J. Phys. Chem. A* **101**, 5111–5116 (1997).

36. Fan, Y., Iwashita, T. & Egami, T. How thermally activated deformation starts in metallic glass. *Nat. Commun.* **5**, 5083 (2014).

37. Berthier, L., Flenner, E., Fullerton, C. J., Scalliet, C. & Singh, M. Efficient swap algorithms for molecular dynamics simulations of equilibrium supercooled liquids. *J. Stat. Mech. Theory Exp.* **2019**, 064004 (2019).

38. Clamp, M. E., Baker, P. G., Stirling, C. J. & Brass, A. Hybrid Monte Carlo: An efficient algorithm for condensed matter simulation. *J. Comput. Chem.* **15**, 838–846 (1994).

39. Mehlig, B., Heermann, D. W. & Forrest, B. M. Hybrid Monte Carlo method for condensed-matter systems. *Phys. Rev. B* **45**, 679–685 (1992).

40. Allen, M. P. & Tildesley, D. J. *Computer simulation of liquids*. (Oxford University Press, 2017).

41. Mendelev, M. I., Sordelet, D. J. & Kramer, M. J. Using atomistic computer simulations to analyze x-ray diffraction data from metallic glasses. *J. Appl. Phys.* **102**, 043501 (2007).

42. Mendelev, M. I. *et al.* Development of suitable interatomic potentials for simulation of liquid and amorphous Cu–Zr alloys. *Philos. Mag.* **89**, 967–987 (2009).

43. Cheng, Y. Q., Ma, E. & Sheng, H. W. Atomic level structure in multicomponent bulk metallic glass. *Phys. Rev. Lett.* **102**, 245501 (2009).

44. Tang, C. & Harrowell, P. Predicting the solid state phase diagram for glass-forming alloys of copper and zirconium. *J. Phys. Condens. Matter* **24**, 245102 (2012).

45. Li, Y., Guo, Q., Kalb, J. A. & Thompson, C. V. Matching glass-forming ability with the density of the amorphous phase. *Science* **322**, 1816–1819 (2008).

46. Zhou, S. H. & Napolitano, R. E. Phase stability for the Cu–Zr system: First-principles, experiments and solution-based modeling. *Acta Mater.* **58**, 2186–2196 (2010).

47. Wang, W. H., Lewandowski, J. J. & Greer, A. L. Understanding the glass-forming ability of $Cu_{50}Zr_{50}$ alloys in terms of a metastable eutectic. *J. Mater. Res.* **20**, 2307–2313 (2005).

48. Gunawardana, K. G. S. H., Wilson, S. R., Mendelev, M. I. & Song, X. Theoretical calculation of the melting curve of Cu-Zr binary alloys. *Phys. Rev. E* **90**, 052403 (2014).

49. Freitas, R., Asta, M. & de Koning, M. Nonequilibrium free-energy calculation of solids using LAMMPS. *Comput. Mater. Sci.* **112**, 333–341 (2016).

50. Pedersen, U. R. Direct calculation of the solid-liquid Gibbs free energy difference in a single equilibrium simulation. *J. Chem. Phys.* **139**, 104102 (2013).

51. Pedersen, U. R., Hummel, F., Kresse, G., Kahl, G. & Dellago, C. Computing Gibbs free energy differences by interface pinning. *Phys. Rev. B* **88**, 094101 (2013).

52. Wessels, V. *et al.* Rapid chemical and topological ordering in supercooled liquid $Cu_{46}Zr_{54}$. *Phys. Rev. B* **83**, 094116 (2011).

53. Wu, X. *et al.* Elucidating the nature of crystallization kinetics in $Zr_{46}Cu_{46}Al_8$ metallic glass through simultaneous WAXS/SAXS measurements. *Appl. Phys. Lett.* **114**, 211903 (2019).

54. Cheng, Y. Q. & Ma, E. Atomic-level structure and structure–property relationship in metallic glasses. *Prog. Mater. Sci.* **56**, 379–473 (2011).

55. Wang, L. *et al.* Low-frequency vibrational modes of stable glasses. *Nat. Commun.* **10**, 26 (2019).

56. Shakerpoor, A., Flenner, E. & Szamel, G. Stability dependence of local structural heterogeneities of stable amorphous solids. *Soft Matter* **16**, 914–920 (2020).

57. Grigera, T. S., Martín-Mayor, V., Parisi, G. & Verrocchio, P. Phonon interpretation of the 'boson peak' in supercooled liquids. *Nature* **422**, 289–292 (2003).

58. Vollmayr, K., Kob, W. & Binder, K. How do the properties of a glass depend on the cooling rate? A computer simulation study of a Lennard‐Jones system. *J. Chem. Phys.* **105**, 4714–4728 (1996).

59. Tkatch, V. I., Limanovskii, A. I., Denisenko, S. N. & Rassolov, S. G. The effect of the melt-spinning processing parameters on the rate of cooling. *Mater. Sci. Eng. A* **323**, 91–96 (2002).

60. Zhu, C., Byrd, R. H., Lu, P. & Nocedal, J. Algorithm 778: L-BFGS-B: Fortran subroutines for large-scale bound-constrained optimization. *ACM Trans. Math. Softw.* **23**, 550–560 (1997).

61. Kresse, G. & Joubert, D. From ultrasoft pseudopotentials to the projector augmented-wave method. *Phys. Rev. B* **59**, 1758–1775 (1999).

62. Blöchl, P. E. Projector augmented-wave method. *Phys. Rev. B* **50**, 17953–17979 (1994).

63. Perdew, J. P., Burke, K. & Ernzerhof, M. Generalized gradient approximation made simple. *Phys. Rev. Lett.* **77**, 3865–3868 (1996).

64. Thompson, A. P. *et al.* LAMMPS - a flexible simulation tool for particle-based materials modeling at the atomic, meso, and continuum scales. *Comput. Phys. Commun.* **271**, 108171 (2022).

65. Nosé, S. A unified formulation of the constant temperature molecular dynamics methods. *J. Chem. Phys.* **81**, 511–519 (1984).

66. Martyna, G. J., Tobias, D. J. & Klein, M. L. Constant pressure molecular dynamics algorithms. *J. Chem. Phys.* **101**, 4177–4189 (1994).

67. Parrinello, M. & Rahman, A. Crystal structure and pair potentials: A molecular-dynamics study. *Phys. Rev. Lett.* **45**, 1196–1199 (1980).

Supplementary information of

# Efficient and accurate simulation of vitrification in multi-component metallic liquids with neural-network potentials

Rui Su[1], Jieyi Yu[1], Pengfei Guan[2,1], Weihua Wang[3]

[1]Institute of Advanced Magnetic Materials, College of Materials & Environmental Engineering, Hangzhou Dianzi University, Hangzhou 310018, P. R. China

[2]Beijing Computational Science Research Center, Beijing 100193, P. R. China

[3]Songshan Lake Materials Laboratory, Dongguan 523808, China

***Corresponding authors：**

R. Su (surui@hdu.edu.cn)

P. F. Guan (pguan@csrc.ac.cn)

**Additional information on the NNP construction and calculated crystalline energies**

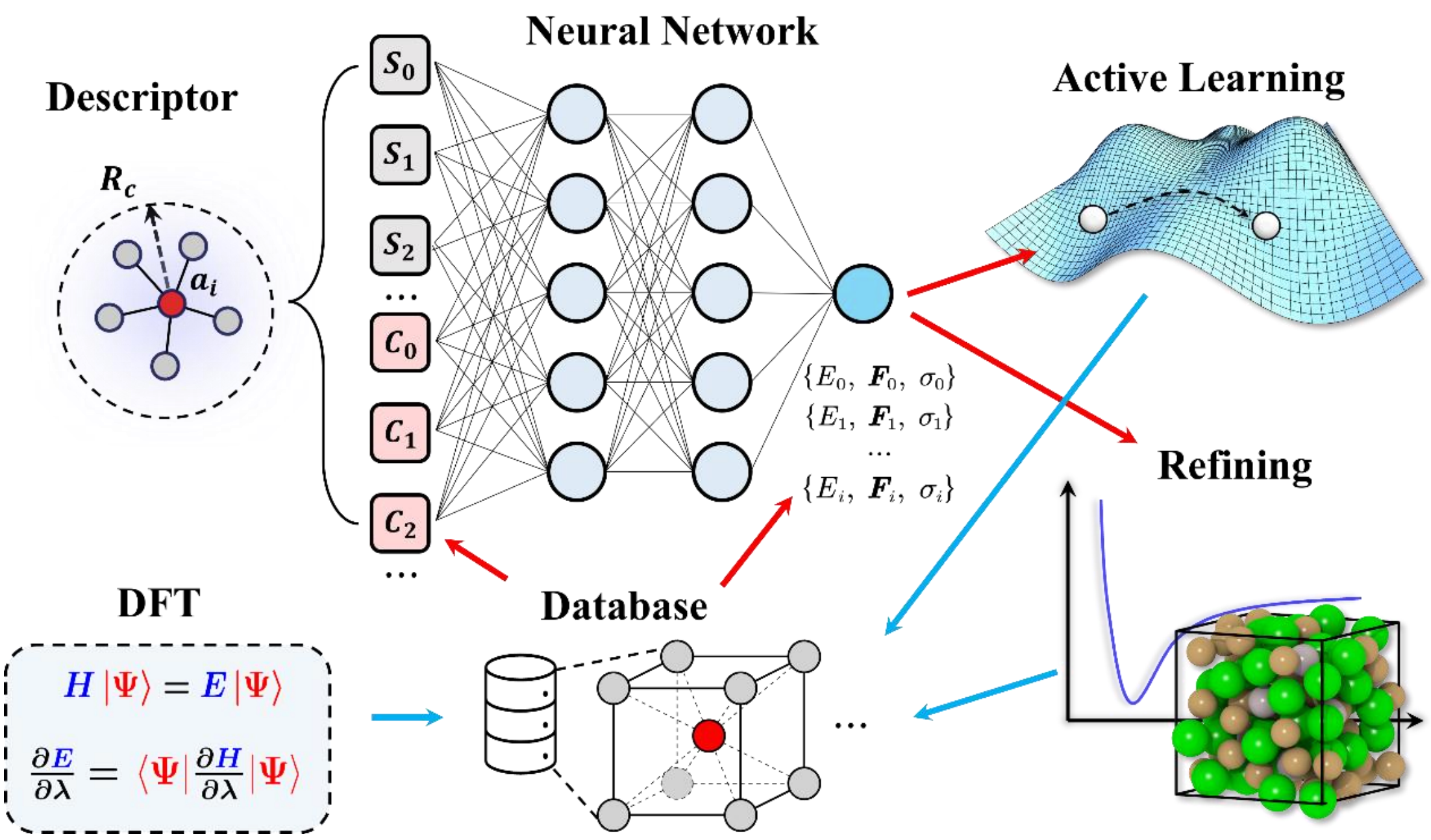

**Figure S1.** Illustration of the NNP construction workflow in NNAP. $S_i$ and $C_i$ refer to the structure-only and the chemical-weighted part of descriptors, respectively.

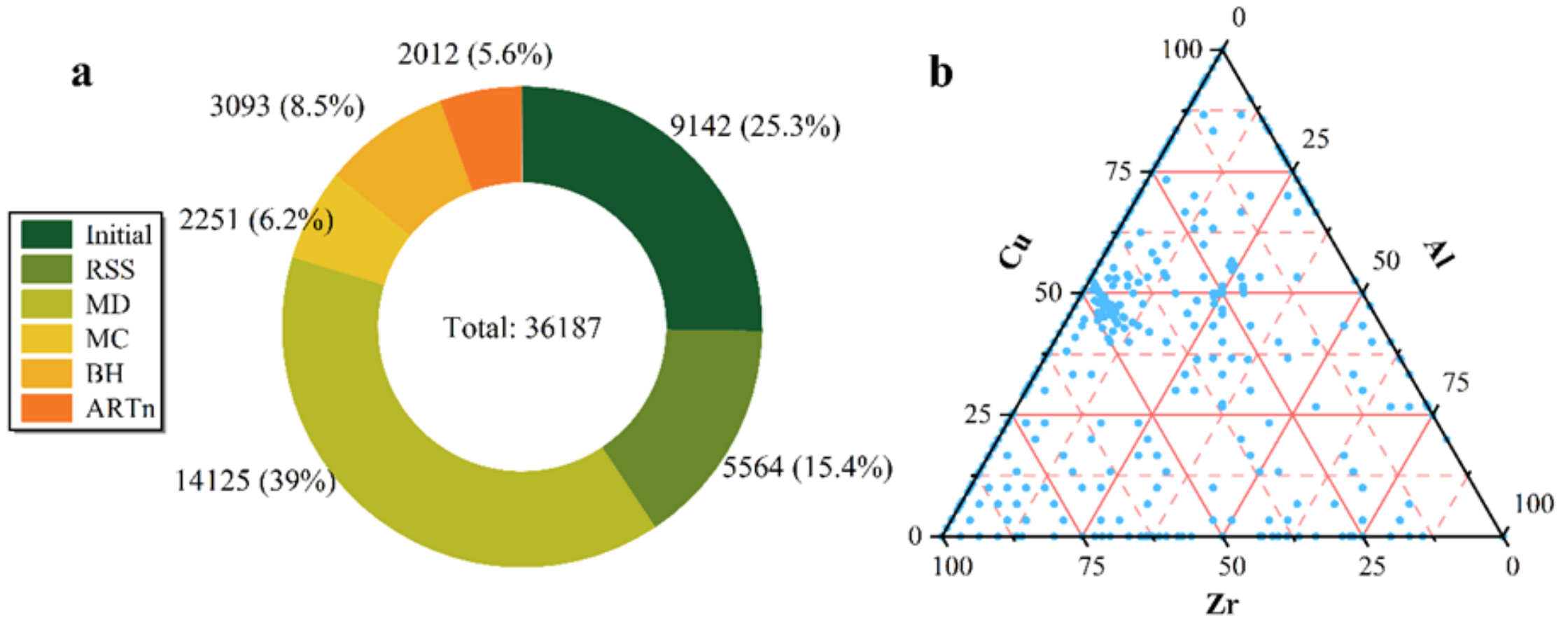

**Figure S2.** Composition of the DFT training database for Zr-Cu-Al systems. (a): Bar plot showing the number of samples with different generating methods; (b): Ternary plot of sampled chemical compositions in the Zr-Cu-Al chemical space.

**Table S1.** Formation enthalpies ($\boldsymbol{\Delta H_f}$)[a] of different crystalline phases.

| Formula | Phase | $\Delta H_{f,0K}$ (eV/atom) | | | | | |
|---|---|---|---|---|---|---|---|
| | | DFT | NNP | CMS | MSZWH | MKOSYP | MSK |
| Zr | *hcp* | 0 | 0 | 0 | 0 | 0 | 0 |
| Zr | *bcc* | 0.084 | 0.0855 | 0.0237 | 0.0517 | 0.0517 | 0.0517 |
| $Zr_2Cu$ | *C11$_b$* | -0.1428 | -0.1453 | -0.1146 | -0.1465 | -0.0935 | -0.0237 |
| ZrCu | *B2* | -0.1052 | -0.1034 | -0.1552 | -0.134 | -0.15 | -0.1391 |
| $Zr_7Cu_{10}$ | $\phi$ | -0.175 | -0.1735 | -0.1074 | -0.1651 | -0.1144 | -0.0949 |
| $ZrCu_2$ | $\sigma$ | -0.129 | -0.1304 | -0.0547 | -0.0857 | -0.0683 | 0.0223 |
| $ZrCu_2$ | *C14* | -0.1172 | -0.1154 | -0.1522 | -0.1664 | -0.1504 | -0.1849 |
| $ZrCu_2$ | *C36* | -0.0903 | -0.0868 | -0.1467 | -0.1611 | -0.1511 | -0.1958 |
| $ZrCu_2$ | *C15* | -0.0662 | -0.0625 | -0.1411 | -0.1345 | -0.1517 | -0.2065 |
| $Zr_3Cu_8$ | $\delta$ | -0.1754 | -0.1769 | -0.1177 | -0.1726 | -0.0893 | -0.0945 |
| $Zr_{14}Cu_{51}$ | $\beta$ | -0.1713 | -0.1778 | -0.0788 | -0.1393 | -0.0868 | -0.0689 |
| $ZrCu_5$ | *C15$_b$* | -0.1266 | -0.1266 | -0.0289 | -0.0989 | -0.034 | -0.0522 |
| Cu | *fcc* | 0 | 0 | 0 | 0 | 0 | 0 |
| **MAE** | - | - | **0.0023** | **0.0631** | **0.0335** | **0.0628** | **0.0898** |
| AlCu | mp-2500[b] | -0.215 | -0.2122 | -0.2458 | - | - | - |
| $Al_2Cu$ | mp-998[b] | -0.1559 | -0.1523 | -0.2467 | - | - | - |
| Al | *fcc* | 0 | 0 | 0 | - | - | - |
| ZrAl | mp-11233[b] | -0.4717 | -0.4783 | -0.4717 | - | - | - |
| $Zr_2Al$ | mp-2557[b] | -0.3607 | -0.3645 | -0.4043 | - | - | - |
| $Zr_3Al$ | mp-1471[b] | -0.3026 | -0.2938 | -0.3118 | - | - | - |
| $ZrAlCu_2$ | mp-3736[b] | -0.3619 | -0.3687 | -0.3211 | - | - | - |
| **MAE** | - | - | **0.0034** | **0.0535** | - | - | - |

(a): $\Delta H_f(\mathrm{Zr_n Cu_m Al_k}) = (E - mE_{\mathrm{Zr}} - nE_{\mathrm{Cu}} - kE_{\mathrm{Al}})/(n+m+k)$

(b): Materials project IDs.

**Active learning algorithm for training configuration selections**

Firstly, we define the ensemble error for a configuration as:

$$\epsilon = \sqrt{\frac{1}{k}\sum_{i}^{k}(E_i - \bar{E})^2} \tag{S1}$$

where $E_i$ and $\bar{E}$ are per-atom potential energy from *i*-th NNP and ensemble average on the *k*-NNP ensemble obtained from last NNP training; secondly, NNP-MD simulations are performed to locate configurations with large ensemble errors employing an online active-learning algorithm:

**Algorithm**: Online active selection of training samples

**Input**: Initial configuration: $\boldsymbol{P_0}$, temperature: $\boldsymbol{T}$, number of MD steps $\boldsymbol{N}$, number of data collection steps $\boldsymbol{N_{collect}}$, max ensemble error: $\boldsymbol{\epsilon_{max}}$, minimal sampling interval: $\boldsymbol{N_{min}}$, sampling threshold: $\boldsymbol{Z_{thresh}}$.

**Output:** Collection of sampled MD configurations: $\boldsymbol{S}$

```
Set S ← {}
Collection of ensemble error: ϵ ← {}
Last MD step of collection: N_last ← 0
for i = 1, 2, ..., N; do
    Take one MD step: P_{i-1} → P_i
    Compute current ensemble error: ϵ_i
    if ϵ_i > ϵ_max:
        stop
    else:
        add ϵ_i to ϵ
    if i > N_collect:
        Compute current sample score: Z_score = (ϵ_i − mean(ϵ))/std(ϵ)
        if Z_score > Z_thresh && i − N_last > N_min:
            Append current configuration P_i to S
            let N_last = i
return S
```

In practice, we perform 4-5 iterations for initial structure at different lattice spacings until stable MD simulations can be performed starting from this structure.

**Performance analysis of the Zr-Cu-Al NNP**

When applying NNP to real-world MD simulations, the major difficulty is the computation complexity due to its complex math forms. In our NNP model, profiling shows that most CPU instructions are spent on the evaluation of the descriptor and their derivatives. In general, the computation complexity of descriptor computation scales linearly as $n_{\max}$ and quadratically as $l_{\max}$. For the current Zr-Cu-Al NNP, we measured the NNP-MD performance as $\sim 0.19$ ms/timestep/core on an Intel Xeon Gold 6140 CPU core. In comparison, we also test the BP-NNP performance as implemented in the AENET package[S1], where the MD performance is $\sim 2.0$ ms/timestep/core with a published BP parameters set[S1] but adjust the cutoff to the same one of our NNP. All testing is performed through the LAMMPS interfaces[S2] to the underlying NNP codes. Since the BP-NNP is trained on the same database with a smaller descriptor size ($N = 72$) and neural network ($72 \times 10 \times 10 \times 1$), our NNP implementation is at least one order of magnitude faster than the previous BP-NNP implementation. However, the EAM potential shows a performance of $\sim 0.002\ ms/timestep/core$, which is about two orders faster than NNP. Fortunately, the NNP parallel scaling is often better than EAM for small-to-medium systems due to the constant communication time, which makes up for some total wall time. In general, we expect the NNP-MD to be $\sim 100$ times slower than the EAM one. Thus, the NNP-MD can only afford quenching rates as low as $10^{10}\ K/s$ for the MD quenching. For example, a $10^{10}\ K/s$ quenching run costs 41 hours with 180 CPU cores for a $10^3$-atom $Zr_{46}Cu_{46}Al_8$ configuration. Such a limitation on the MG sample stability hinders the preparation of stable MG samples, which are critical to understanding the structure and mechanical properties evolutions during the vitrification of MGs. This is the primary motivation for the development of the NNP-SHMC, which dramatically accelerates the equilibrium process of deeply supercooled liquid and the vitrification process of MGs.

**Test on the transferability of Zr-Cu-Al NNP**

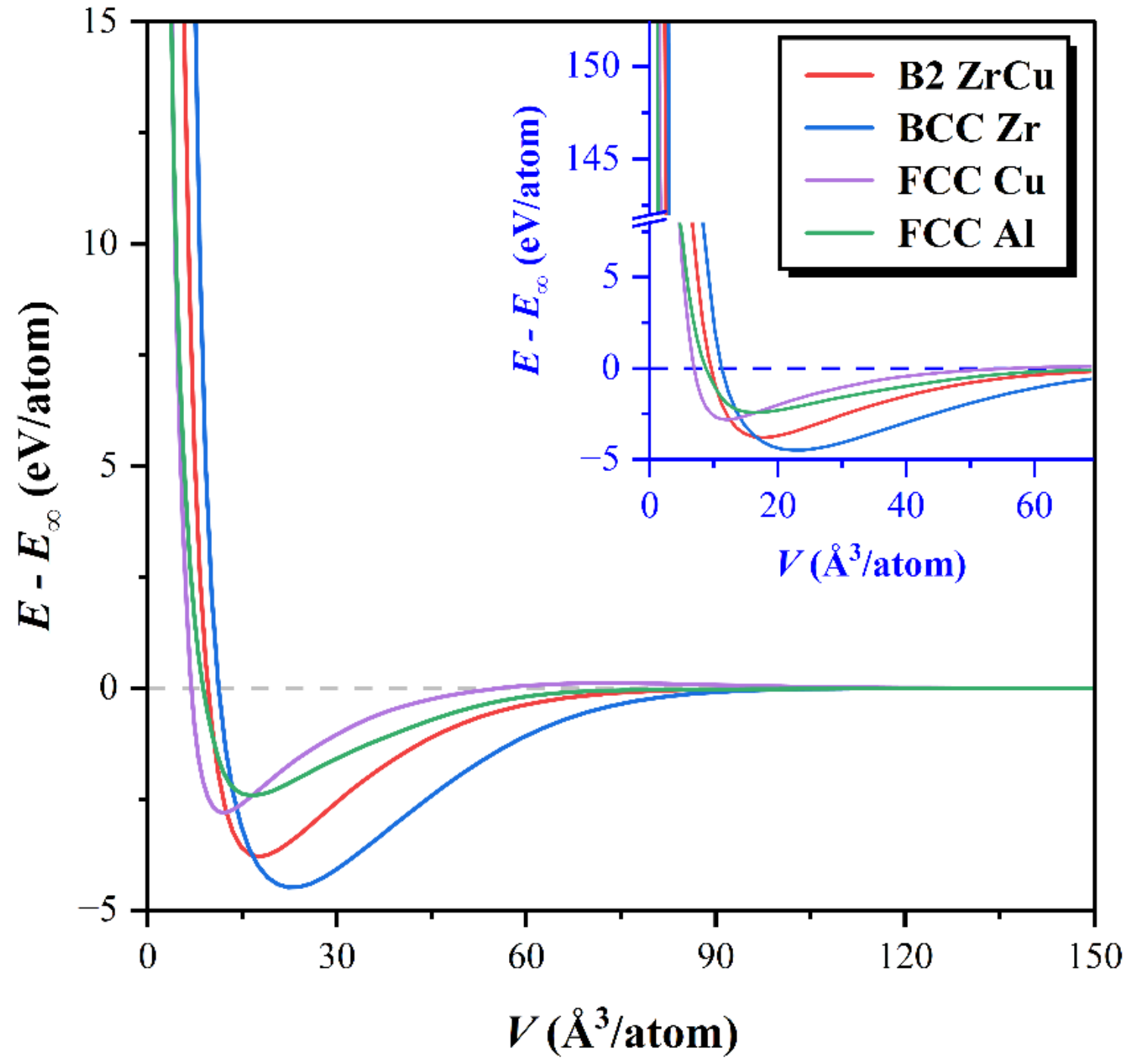

**Figure S3.** Energy-volume curves in the full range of crystal volumes for B2-ZrCu (solid red lines), bcc-Zr (solid blue lines), fcc-Cu (solid purple lines), and fcc-Al (solid green lines). The insect plot shows $E$-$V$ curves in the extended energy range.

Transferability of the NNPs is critical for general MD simulation purposes and has been under debate since its formulation does not incorporate physical information, which might result unphysical behaviors under some limiting situations. While a thoroughly exploration of the transferability of even a classical potential is hard, energy-volume curves would provide insightful results on the potential behavior in a large range of atom separations. Pun *et al.* have shown that some AI-NNPs show unphysical energy-volume curves when the crystals are highly compressed, and they suggest a physically informed NNP formalism that solves such a problem[S3]. While their method offers fresh insights into this problem, we found that our NNP does not suffer from such disadvantages either. As shown in **Fig. S3**, the rigid $E$-$V$ curves for four crystalline phases are plotted against the per-atom volume in a large range of $0$ to $150$ Å$^3$. It can be clearly seen that the energy curves are very smooth and the divergence of energy near $V \to 0$ is obvious. Thus, our NNP should be expected to have the same level of transferability as the physical potentials like EAMs or the physically informed NNP by Pun *et. al*. We further note that the similar behavior of $E$-$V$ curves has been observed in many other tested systems of us, which shows that it is a general feature of our NNP formalism.

**Optimization of the SHMC parameters**

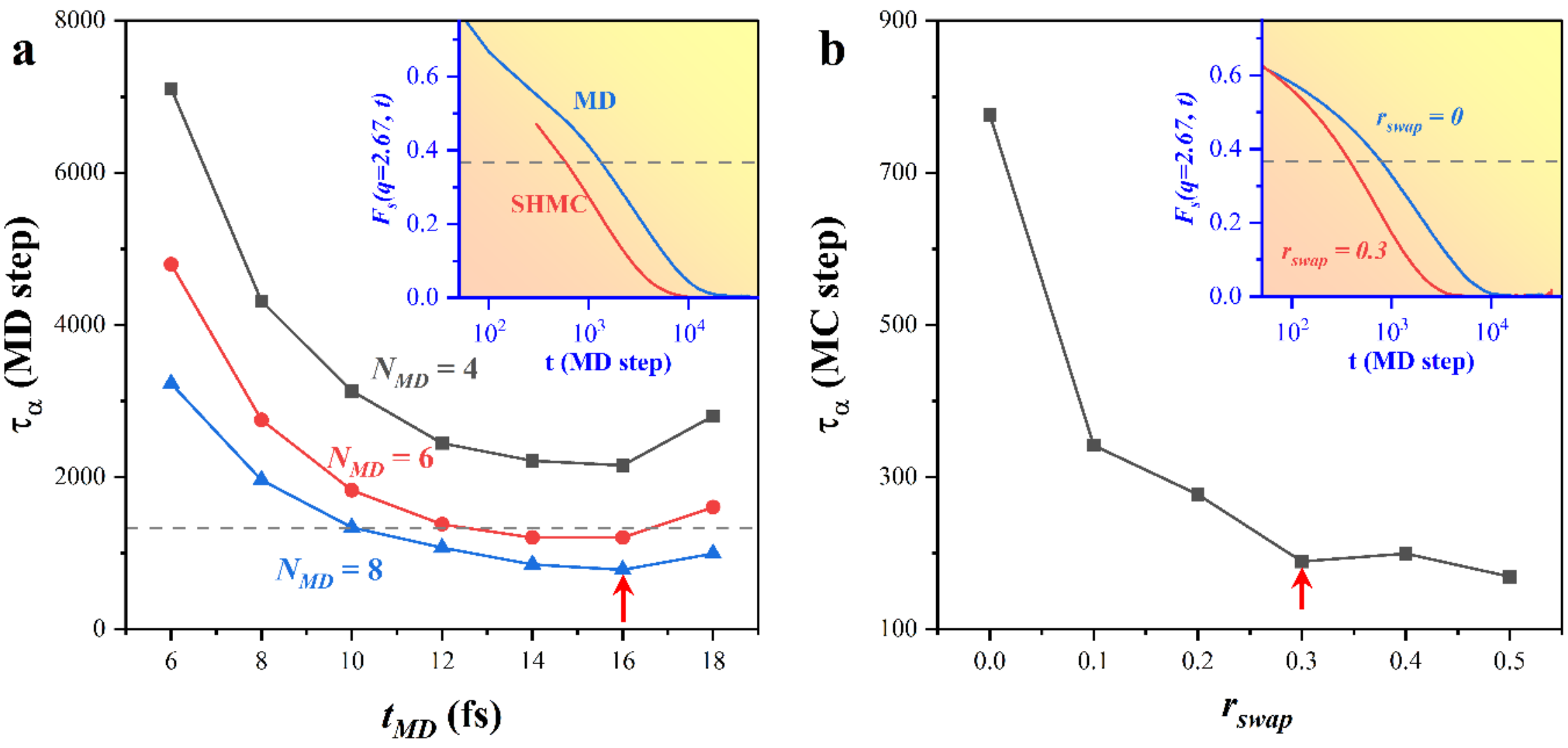

**Figure S4.** Optimization of the SHMC parameters. (a): Optimization of $\boldsymbol{t_{MD}}$ and $\boldsymbol{N_{MD}}$. (b): Optimization of $\boldsymbol{r_{swap}}$. Intersect plots show SISF plots of different parameter sets, respectively. The optimized SHMC parameter sets are labeled by red arrows. All $\boldsymbol{\tau_\alpha}$ values are calculated in the unit of the MD integration step for fair comparison between MD and SHMC results.

To achieve the best acceleration ratio, we optimize the combination of three primary parameters: $N_{MD}$, $t_{MD}$ and $r_{swap}$ of SHMC simulations. **Figure S4** shows our optimization process for three SHMC parameters: $N_{MD}$, $t_{MD}$, and $r_{swap}$ at an arbitrarily selected thermodynamic equilibrium temperature. Firstly, we optimized the MD parameters: $N_{MD}$, $t_{MD}$ by fixing $r_{swap} = 0$, and the results are shown in **Fig. S4a**. We find that $t_{MD} = 16\,fs$ approximately corresponds to the minimum structural relaxation time, which is independent of the choice of the $N_{MD}$ values. Meanwhile, the relaxation time continuously decreases as $N_{MD}$ increases from 4 to 8. However, a further increase of $N_{MD}$ brings only a slightly reduce of $\tau_\alpha$ at the price of significantly increased computing cost of each MC sweep. Thus, $N_{MD} = 8$ and $t_{MD} = 16\,fs$ are selected for our following simulations, resulting in a $\tau_\alpha$ smaller than the MD one.

We further optimize the ratio of swap attempts $r_{swap}$ that determines the number of swap attempts in each MC sweep as: $N_{swap} = r_{swap} \times N_{atoms}$. As shown in **Fig. S4b**, the decrease of $\tau_\alpha$ reaches saturation as $r_{swap}$ is increased to 0.3. For the arbitrarily selected temperature $T = 950\,K$ and $r_{swap} = 0.3$, the $\tau_\alpha$ of SHMC is four times faster than the standard MD simulation. Further tests show that although the acceleration ratio between SHMC and MD simulations is temperature-dependent, the optimal $r_{swap}$ is insensitive to the selected temperature. Thus, we choose the fixed parameter set: $N_{MD} = 8$, $t_{MD} = 16\,fs$, and $r_{swap} = 0.3$ for all the following simulations.

**Calculation of the free energies of crystalline phases**

We calculate the Gibbs free energies at $P = 0\,GPa$ at different temperatures by the Frenkel-Ladd method in the non-equilibrium thermodynamical integration (TI) approach[S4]. In this approach, the system Helmholtz free energy is calculated as:

$$F_0(N,V,T) = F_E(N,V,T) + \Delta F(N,V,T) \tag{S2}$$

where $F_E(N,V,T)$ is the free energy of the referencing Einstein solid and the free energy difference $\Delta F(N,V,T)$ between current and reference Hamiltonian is calculated as:

$$\Delta F = \int_0^1 d\lambda \left\langle \frac{\partial H}{\partial \lambda} \right\rangle_\lambda = \frac{1}{2}\left(W_{0\to E}^{irr} - W_{E\to 0}^{irr}\right) \tag{S3}$$

where $\langle \cdots \rangle$ denotes for ensemble averaging, $0$ and $E$ stand for initial NNP Hamiltonian and final Einstein solid one, respectively.

In the TI calculation, $\langle \partial H / \partial \lambda \rangle_\lambda$ is calculated as the system Hamiltonian changes from 0 to $E$ by varying the time-dependent parameter $\lambda$:

$$H(\lambda) = \lambda H(0) + (1-\lambda) H(E) \tag{S4}$$

where $\lambda$ changes from 0 to 1 in the switching time $t_s$. In this work, the switching function: $\lambda(t) = t^5(70t^4 - 315t^3 + 540t^2 - 420t + 126)$ as suggested by Freitas *et al.*[S4] is used.

The irreversible work along the TI path is calculated as:

$$W_{0\to E}^{irr} = \int_0^{t_s} dt \frac{d\lambda}{dt} \left( \frac{\partial H}{\partial \lambda} \right)_{\Gamma(t)} \tag{S5}$$

where $\Gamma(t)$ denotes the phase-space trajectory produced by the simulation. In the non-equilibrium TI, the reversible part of work (free energy difference) is estimated by canceling the irreversible part as:

$$\Delta F = \frac{1}{2}\left(W_{0\to E}^{irr} - W_{E\to 0}^{irr}\right) \tag{S6}$$

The absolute free energy of the Einstein solid is:

$$F_E(N,V,T) = 3Nk_B T \ln\left(\frac{\hbar\omega}{k_B T}\right) \tag{S7}$$

where the oscillator frequency $\omega$ is obtained from spring constant: $k = m\omega^2$. The mean squared displacement $\langle \Delta r^2 \rangle$ is used to calculate the spring constant $k$:

$$k = \frac{3k_B T}{\langle \Delta r^2 \rangle} \quad (S8)$$

To obtain Gibbs free energy, the system is first equilibrated at external pressure $P$. Then the Helmholtz free energy $F$ is evaluated at fixed volume. The Gibbs free energy is obtained as $G = F + PV$.

In this work, we calculate the free energies of crystalline phases as follows: firstly, the system is equilibrated for $50\,ps$ at a given temperature and pressure in a *NpT* ensemble and the spring constant $k$ is estimated; secondly, we fix the box and vary $\lambda$ to perform forward and backward TI runs to calculate the free energy difference $\Delta F$. Finally, Gibbs free energy is obtained by adding the contribution from Einstein solid. For both forward and backward TI runs, we first equilibrate the system for $t_{eq} = 10\,ps$ and switch $\lambda$ with a switching time $t_s = 50\,ps$. Supercells containing 2,000, 1,800, and 3,264 atoms are used for ZrCu, $Zr_2Cu$, and $Zr_7Cu_{10}$, respectively.

**Calculation of the melting temperature of B2-ZrCu**

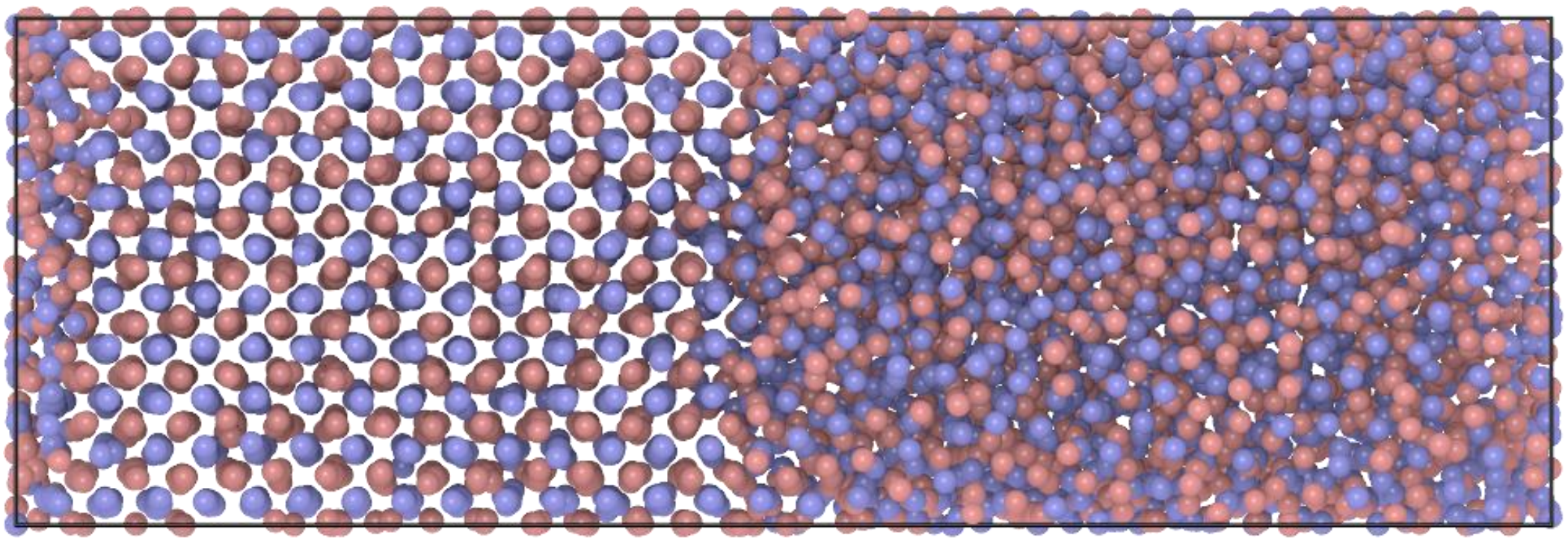

**Figure S5.** MD snapshot of the equilibrated solid-liquid interface of B2-ZrCu. The Cu and Zr atoms are shown as red and blue atoms, respectively.

The melting temperature of B2-ZrCu is calculated by the "interface pinning" (IP) method[S5] at $P = 0\,GPa$. IP method holds the solid-liquid interface at $T$ by adding a harmonic bias term to the potential energy:

$$U'(\mathbf{R}) = U(\mathbf{R}) + \frac{\kappa}{2}(Q(\mathbf{R}) - a)^2 \quad (S9)$$

where $Q(\mathbf{R})$ is a global order parameter that depends on the number of crystalline atoms. $\kappa$ is the harmonic sprint constant. $a$ corresponds to the "interface position" in the 1D space of order parameter $Q(\mathbf{R})$.

The chemical potential difference between solid and liquid phases is calculated as:

$$\Delta\mu = -\kappa(\langle Q \rangle' - a)\Delta Q / N \quad (S10)$$

The entropy difference $\Delta s$ can be calculated from basic thermodynamic relation:

$$\Delta s = \frac{\Delta u + p\Delta v - \Delta\mu}{T} \tag{S11}$$

where $\Delta\mu$ and $\Delta v$ are differences in internal energy and specific volume between crystalline and liquid phases, respectively. From a starting guess $T_0$, the melting temperature $T$ estimation is made through Newton's iteration:

$$T^{i+1} = T^i + \frac{\Delta\mu}{\Delta s} \tag{S12}$$

The iteration stopped when $T$ variance was less than statistical error.

In this work, we use the translation order parameter[S2] $\rho_k$ as $Q(\mathbf{R})$:

$$Q(R) \equiv |\rho_k| = \left| N^{-\frac{1}{2}} \sum_{j=1}^{N} \exp(-i\mathbf{k} \cdot \mathbf{r}_j) \right| \tag{S13}$$

For the B2-CuZr, we construct a $8 \times 8 \times 20$ supercell elongating along the $z$ axis. $\rho_k$ is evaluated using the wave vector $\mathbf{k} = (2\pi n_x/X, 0,0)$ where $n_x = 16$. The biased harmonic potential is applied using $\kappa = 4.0$ and $a = 26$. The solid-liquid interface is equilibrated using a $Np_zT$ ensemble for $4\,ns$ and the last $200\,ps$ is used to calculate the ensemble-averaged $\langle Q \rangle'$. **Figure S5** shows a snapshot of the equilibrated solid-liquid interface at $T_m$ using the NNP potential.

# References

S1. Artrith, N. & Urban, A. An implementation of artificial neural-network potentials for atomistic materials simulations: Performance for TiO2. *Computational Materials Science* **114**, 135–150 (2016).

S2. Thompson, A. P. *et al.* LAMMPS - a flexible simulation tool for particle-based materials modeling at the atomic, meso, and continuum scales. *Computer Physics Communications* **271**, 108171 (2022).

S3. Pun, G. P. P., Batra, R., Ramprasad, R. & Mishin, Y. Physically informed artificial neural networks for atomistic modeling of materials. *Nature Communications* **10**, 1–10 (2019).

S4. Freitas, R., Asta, M. & de Koning, M. Nonequilibrium free-energy calculation of solids using LAMMPS. *Comp. Mater. Sci.* **112**, 333–341 (2016).

S5. Pedersen, U. R. Direct calculation of the solid-liquid Gibbs free energy difference in a single equilibrium simulation. *J. Chem. Phys.* **139**, 104102 (2013).